\documentclass[pra,twocolumn,groupedaddress,showpacs,floatfix]{revtex4}
\usepackage{amsmath,amssymb,amsfonts}
\usepackage{mathcomp}
\usepackage{comment}
\usepackage{graphicx,subfigure}
\usepackage{txfonts}
\usepackage{epstopdf}
\usepackage[title]{appendix}

\usepackage[usenames, dvipsnames]{color}
\usepackage{soul}
\usepackage[normalem]{ulem}

\usepackage{bbold}
\usepackage{footnote}

\usepackage{xr-hyper} 
\usepackage[unicode]{hyperref} 
\hypersetup{
    unicode=true,                                           
    plainpages=false,
    pdffitwindow=true,                                      
    colorlinks=true,                                        
    linkcolor=blue,                                        
    citecolor=blue,                                         
    filecolor=blue,                                        
    urlcolor=blue                                          
}

\newcommand{\bra}[1]{\langle #1\vert}
\newcommand{\lbra}[1]{\langle \mkern-3mu \langle #1\vert}
\newcommand{\ket}[1]{\vert #1\rangle}
\newcommand{\lket}[1]{\vert #1 \rangle \mkern-3mu \rangle}

\newcommand{\ev}[1]{\langle #1 \rangle}
\newcommand{\pd}[1]{\frac{\partial}{\partial #1}}

\newcommand{\refeq}[1]{Eq.~(\ref{#1})}

\newcommand{\reffig}[1]{Fig.~\ref{#1}}

\newcommand*\colvec[3][]{
  
   \begin{pmatrix}\ifx\relax#1\relax\else#1\\\fi#2\\#3\end{pmatrix}
}
 \newcommand{\ldts}[0]{\mathinner{{\ldotp}{\ldotp}{\ldotp}}}

\newcommand{\cm}{$\textrm{cm}$}
\newcommand{\ns}{$\textrm{ns}$}
\newcommand{\ps}{$\textrm{ps}$}

\newcommand{\drangle}{\rangle \mkern-3mu\rangle}

\begin{document}
\title{Perturbation approach for computing frequency- and time-resolved photon correlation functions} 
\author{David I. H. Holdaway}
\affiliation{Department of Physics and Astronomy, University College London, Gower Street, WC1E 6BT, London, United Kingdom}
\author{Valentina Notararigo}
\affiliation{Department of Physics and Astronomy, University College London, Gower Street, WC1E 6BT, London, United Kingdom}
\author{Alexandra Olaya-Castro}
\email{a.olaya@ucl.ac.uk}
\affiliation{Department of Physics and Astronomy, University College London, Gower Street, WC1E 6BT, London, United Kingdom}


 \begin{abstract}
We propose an alternative formulation of the sensor method presented in [Phys. Rev. Lett 109, 183601 (2012)] for the calculation of frequency-filtered and time-resolved photon correlations. Our approach is based on an algebraic expansion of the joint steady state of quantum emitter and sensors with respect to the emitter-sensor coupling  parameter $\epsilon$. This allows us to express photon correlations in terms of the open quantum dynamics of the emitting system only and ensures that computation of correlations are independent on the choice of a small value of $\epsilon$.  Moreover, using time-dependent perturbation theory, we are able to express the frequency- and time-resolved second-order photon correlation as the addition of three components, each of which gives insight into the physical processes dominating the correlation at different time scales. We consider a bio-inspired vibronic dimer model to illustrate the agreement between the original formulation and our approach.
 \end{abstract}


\maketitle

\section{Introduction}
Single-photon coincidence measurements have been recognised as a fundamental theoretical and experimental methodology to characterize quantum properties both of light \cite{Glauber1963, Glauber2006, Grangier1986, Lounis2000} as well as those of the emitting source \cite{Olaya2001, Michler2000, Moreau2001}. Particular focus has been placed on investigation of the second-order photon correlation function as the lowest order of correlations capable of probing non-classical phenomena \cite{Glauber2006}. Formally, such normally-ordered two-photon correlation function is defined as \cite{VogelBook}
\begin{equation}
g^{(2)}(t_1,t_2) =\frac{\ev{\mathcal{T}_{-}[\hat{A}^{\dagger}(t_1)\hat{A}^{\dagger}(t_2)]\mathcal{T}_{+}[\hat{A}(t_2)\hat{A}(t_1) ]}} {\ev{\hat{A}^{\dagger}(t_1)\hat{A}(t_1)}\ev{\hat{A}^{\dagger}(t_2)\hat{A}(t_2)}} \;,
\label{eq:g2def}
\end{equation}
with $\hat{A}$ being the field operator and  $\mathcal{T}_-$ and $\mathcal{T}_+$ the time-ordering and antiordering superoperators necessary for a consistent physical description \cite{VogelBook}.  Here $\mathcal{T}_-$ increases time arguments to the right in products of creation operators, while $\mathcal{T}_+$ increases time arguments to the left in products of annihilation operators. 

In the context of photon counting experiments it has also become clearer that spectral filtering of optical signals --and its associated trade off between frequency and time resolution-- opens up the door for the investigation of  variety of phenomena in quantum optics \cite{VogelBook, Sallen2010, Peiris2015, Grunwald2015, Silva2016}. The energy-time Fourier uncertainty relation imposes a constraint on the precision with which arrival time and frequency of a photon can be measured \cite{Eberly1977,Brenner1982}. Rather than being a limitation, this uncertainty has shown to offer a potential for novel investigations of quantum phenomena ranging from the identification and manipulation of new types of photon quantum correlations \cite{DelValle2012,DelValleErratum, Tudela2013, Tudela2015,  Peiris2015} to the development of new protocols for the preparation and readout of entangled photons \cite{Flayac2014,PRL2017_Peiris}. It has also recently been discussed how frequency- and time-resolved photon correlation measurements can provide insights into the emitter dynamics which are complementary to the information provided by coherent multi-dimensional spectroscopy \cite{Brixner2005} --the later being the ultrafast, non-linear  technique capable of probing of quantum coherence dynamics in a variety of biomolecular and chemical systems (for a review see \cite{Scholes2017}).
 
Filter-dependent correlation functions are defined in terms of filtered emission operators $\hat{A}_F(t) = \int_0^{\infty} F(t') \hat{A}(t-t') dt'$ and $\hat{A}^{\dagger}_F(t) = \int_0^{\infty} F(t') \hat{A}^{\dagger}(t-t') dt'$ with $F(t)$ the one sided Fourier transform of the frequency filter function. 
The filtered two-time correlation function can be written as $g^{(2)}_{F_1,F_2}(T_1,T_2)$ and is defined identically to \refeq{eq:g2def}, but with the substitutions $\hat{A}^{(\dagger)}(t_1) \rightarrow \hat{A}^{(\dagger)}_{F_1}(T_1)$ and $\hat{A}^{(\dagger)}(t_2) \rightarrow \hat{A}^{(\dagger)}_{F_2}(T_2)$ with $F_j(t)$ the time and space filter functions for each detector \cite{VogelBook, Knoll1986, Cresser1987, Bel2009, Kamide2015, PRA16_Kilin}.
Due to the convoluted definition of $\hat{A}_{F_j}^{(\dagger)}$, calculating $g^{(2)}_{F_1,F_2}(T_1,T_2)$ involves computing a four dimensional integral with the time ordering applying within this set of integrals, thereby making such a calculation non-trivial. Higher-order correlations $g^{(n)}_{F_1\ldots F_n}(T_1 \ldots T_n)$ are defined in a similar way, although their theoretical computation becomes more difficult. Thus, a full theoretical understanding of the effects of such filters in the photon statistics has only recently been possible with the development of methods that can overcome the computational complexity \cite{DelValle2012,DelValleErratum}.

In particular, Ref.\cite{DelValle2012} has put forward an  efficient sensor method for calculating these frequency and time-resolved correlation functions, which avoids the need to explicitly compute the multidimensional integral set. The methodology involves accounting, in a quantum mechanical manner, for a weak coupling between the quantum emitter and a set of sensors, each of which is represented as a two-level system. In the limit of vanishing system-sensor coupling, the sensor population correlations are shown to quantify the photon correlations of interest. As originally proposed, this method relies on the explicit use of a numerically small value for the system-sensor coupling. Hence, accurate determination of the photon-correlations function demands testing for convergence. The need for a small parameter may also lead to numerical calculations exhibiting  instabilities. Moreover, the method requires solving the quantum dynamics of the joint emitter-sensors state and therefore the dimensionality of the density matrix can become a problem for quantum systems of large dimensions. 

In this paper, we report an alternative formulation of the sensor method that addresses the above issues. We propose a formalism that allows us to express photon correlations fully in terms of the quantum dynamics of the emitting system while at the same time eliminating the dependence on a specific value for the small parameter. In our formalism the small parameter vanishes algebraically in all the expressions for multi-photon correlations. Furthermore, using time-dependent perturbation theory, we are able to express the second-order photon correlation as the addition of three components, which provide insight into the physical processes dominating the emission at different time scales.

The paper is organised as follows: Sec.~\ref{sec:derivation} summarises the original presentation of the sensor method and motivates the development of an alternative formulation.  Sec.~\ref{sec:correlation at zero time delay} presents the approach to derive the steady state system and photon correlations at zero-time delay. Sec.~\ref{sec:tdep_pert} explains the derivation of time-dependent correlation functions for finite detection delays, Sec.~\ref{sec:numerics} illustrates the agreement between our approach and the original sensor method  for a bio-inspired vibronic dimer model, and Sec.~\ref{sec:conc} concludes. 

\section{Motivation \label{sec:derivation}}
\begin{figure}[t]
	\centering
	\includegraphics[width=0.5\textwidth]{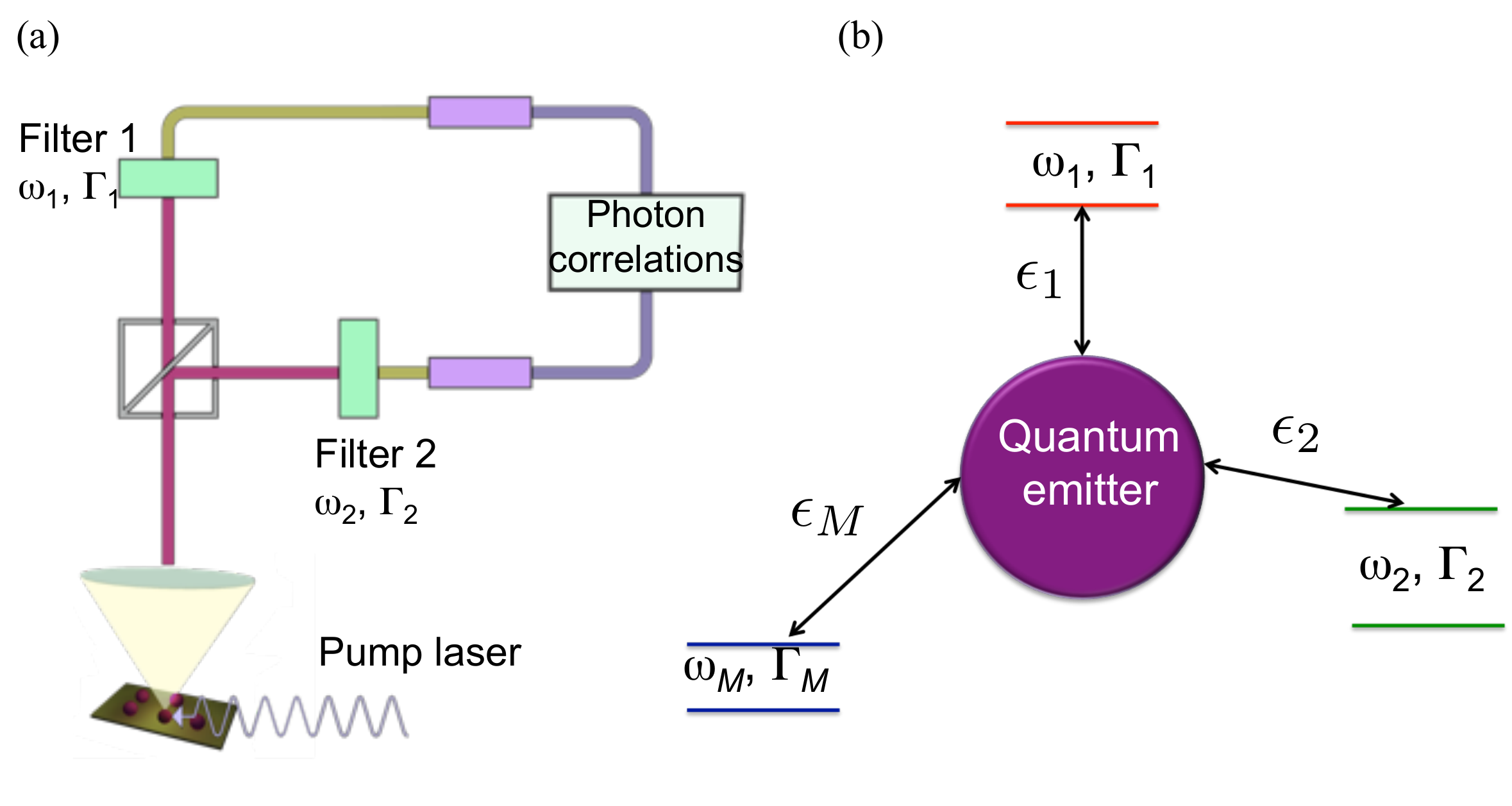} 
	\caption{(Color online) (a) Schematic  of a experimental setup to measure frequency-resolved photon correlations similar to the one used in Ref. \cite{PRL2017_Peiris}) . (b) Diagram of the sensor method proposed in Ref. \cite{DelValle2012} to compute frequency-filtered correlations. Each sensor, represented by a two-level system, is quantum mechanically coupled  to the quantum emitter with coupling strength $\epsilon_m$ with $m=1,\dots, M$. Photon correlations are given by normally-ordered sensor operator correlations in the limit when $\epsilon_1,\dots,\epsilon_M\rightarrow 0$} .
	\label{Fig1}
\end{figure}

As proposed in \cite{DelValle2012}, the sensor method for calculating the $M-$photon correlation function involves simulating the dynamics of a quantum emitter with Hamiltonian  $H_0$, weakly coupled to $M$ sensors represented by two-level systems, labelled with $m=1,\cdots, M$, and ground and excited states $|0_m\rangle$ and $|1_m\rangle$, respectively. Each sensor has an associated Hamiltonian $H_m=\omega_m \varsigma_m^{\dag} \varsigma_m$ with annihilation operator $\varsigma_m=|0_m \rangle \langle 1_m|$ and transition frequency $\omega_m$ set to match the emission frequency to be measured. The interaction Hamiltonian between the quantum emitter and the $m-$th sensor is given by
$H_{e,m}=\epsilon_m \big ( {a}_m \varsigma_m^{\dagger}+{a}_m^{\dagger} \varsigma_m \big )$,
with the coupling strength $\epsilon_m$ being small enough to neglect back action. For generality, we have considered that the emission operators $a_j$ coupled to each sensor can be different. This is the case whenever local resolution is achievable in a multipartite quantum emitter or when emitting transitions can be distinguished via fluorescence polarization detection as it happens, for instance, in single light-harvesting complexes \cite{JPC2011_Pulleritis}. In such a scenario, the frequency filters illustrated in the envisioned experimental setup (see \reffig{Fig1}(a)) will also be polarizing filters.

Considering Markovian relaxation channels for both the emitter and the sensors, the joint emitter-sensors density matrix $\hat{\rho}$ satisfies the master equation $\pd{t} \hat{\rho} = \mathcal{L}(\hat{\rho})$, with the Liouvillian conveniently split as ($\hbar=1$)
\begin{equation}\label{eq:L}
\mathcal{L}(\hat{\rho}) = \mathcal{L}_0(\hat{\rho}) + \sum_{m=1}^{M} \big ( \mathcal{L}_{m}(\hat{\rho}) -i \, [H_{e,m},\hat{\rho}]  \big ),
\end{equation}
with
\begin{align} 
	\mathcal{L}_0(\hat{\rho}) &= 
	-i \, [H_0,\hat{\rho}] + \sum_i \frac{1}{2}\gamma_{c_i} \mathcal{L}_{c_i}(\hat{\rho}),  \label{eq:L0}  \\
	\mathcal{L}_{m}(\hat{\rho}) &=   -i \, [H_m,\hat{\rho}] + \frac{1}{2}\Gamma_m \mathcal{L}_{\varsigma_m}(\hat{\rho}) \label{eq:Lm} . 
\end{align}
The superoperators on the right hand side of \refeq{eq:L0} have the Lindblad form i.e. $\mathcal{L}_{c_i}(\hat{\rho})=2c_i\hat{\rho}c_i^{\dag}-c_i^{\dag}c_i\hat{\rho}-\hat{\rho}c_i^{\dag}c_i$ for a system jump operator $c_i$ and a relaxation process at rate $\gamma_{c_i}$. Same holds for $\mathcal{L}_{\varsigma_m}$ in \refeq{eq:Lm} describing the decay of the $m$th sensor with jump operator $\varsigma_m$ at a rate $\Gamma_m$. 
%
%
In the limit of $\epsilon_m$ satisfying $\epsilon_m\ll\sqrt{\Gamma \gamma_{Q}/2}$ with $\gamma_Q$ the smallest transition rate within the emitter dynamics, and sensor populations satisfying  $\langle n_m \rangle = \langle \varsigma_m^{\dagger} \varsigma_m \rangle \ll 1$,   intensity-intensity correlations of the form $\langle :n_1 n_2 \ldots n_M:\rangle$  are directly related to the $M$th order photon correlation functions \cite{DelValle2012,DelValleErratum}:
\begin{equation}\label{eq:corr_func}
	g_{\Gamma_1...\Gamma_M}^{(M)}(\omega_1,T_1;...;\omega_M,T_M)= 
	\lim_{\epsilon_1,...,\epsilon_M \to 0} \,\frac{\langle :n_1(T_1)...n_M(T_M):\rangle}
	{\langle n_1(T_1)\rangle \, ... \, \langle n_M(T_M)\rangle}
\end{equation}
{\small
\begin{equation}
\label{eq:n_i}
\langle :n_1(T_1)...n_M(T_M):\rangle = \, \frac{\epsilon_1^2...\epsilon_M^2}
	{\Gamma_1...\Gamma_M} \: (2\pi)^M \: S_{\Gamma_1...\Gamma_M}^{(M)}
	(\omega_1,T_1;...;\omega_M,T_M)
\end{equation}}
with~\cite{VogelBook}
\begin{equation}\begin{split}\label{eq:g_Morder}
	&S_{\Gamma_1...\Gamma_M}^{(M)} ( \omega_1,\cdots,\omega_M; T_1,\cdots ,T_M) =  \\
	& = \int_{-\infty}^{\infty} \!\!\! dt_1' \int_{-\infty}^{\infty} \!\!\! dt_{M+1}' \;\; 
	F_1^*(T_1-t_1') \;F_1(T_1-t_{M+1}')\\
	& \quad ... \int_{-\infty}^{\infty} \!\!\! dt_M' \int_{-\infty}^{\infty} \!\!\! dt_{2M}' \;\; 
	F_M^*(T_M-t_M') \;F_M(T_M-t_{2M}')\\
	& \quad \quad \langle \mathcal{T}_- \big[a_{1}^{(+)}(t_1') \, ... \, 
	a_{M}^{(+)}(t_M') \big]
	\: \mathcal{T}_+ \big[a_{1}^{(-)}(t_{M+1}') \, ... \, a_{M}^{(-)}(t_{2M}') \big] \rangle.
\end{split}\end{equation}
The filter functions correspond to a Cauchy--Lorentz distribution i.e. $F_m(t) = \theta(t) \, \Gamma_m/2 \, \exp[-(\Gamma_m/2 +i\omega_m)t]$ with $\theta(t)$ the Heaviside function, and can be realised, for instance, via a Fabry-Perot interferometer when the reflection coefficient tends to unity  \cite{Eberly1977}. The experimental setup we envision is sketched in \reffig{Fig1}(a) and the theoretical calculation of frequency-filtered photon correlations through the sensor method is illustrated in \reffig{Fig1}(b).

The original presentation of \refeq{eq:corr_func}  in \cite{DelValle2012} omitted the normal order. Without the normal order this function yields unphysical results for a finite delay time. In an  Erratum \cite{DelValleErratum} the authors clarified that normal order is implied through the proof of \refeq{eq:corr_func}, though it turns out to be unnecessary for zero time delay. Since \refeq{eq:corr_func} is the departing point of our work, we have carried out a consistency check of its proof as discussed in Appendix \ref{app:proof}.

The method as proposed in \cite{DelValle2012} is conceptually clear yet, in practice, its computation involves tackling some numerical challenges. Assuming that all the sensor couplings are identical,  $\epsilon_j=\epsilon$, the numerical calculations of photon correlations rely on the choice of a system-sensor coupling $\epsilon$ that is numerically small, but not {\it so} small that adding or subtracting terms of order $\epsilon^{2M}$ to or from terms of order $\epsilon^0$ causes problems within double precision arithmetic. The procedure then involves checking convergence and stability of the numerical results for different values of $\epsilon$. Most importantly, computation of photon correlations at zero-time delay requires to numerically finding the zero eigenvalue of the Liouvillian superoperator associated to the joint emitter system plus sensors. This means that computing $g_{\Gamma_1 \ldts \Gamma_M}^{(M)}(\omega_1,T;\ldts;\omega_M,T)$ when $T\rightarrow \infty$, involves calculating the eigenvector with a zero eigenvalue of a matrix $4^{M}$ times larger than that of the quantum emitter alone \cite{Tudela2013}. Similarly, for time-resolved correlations, the calculation involves time propagation in the joint state space of the system and sensors. Evidently, as the dimensionality of the system is larger these numerical challenges become more demanding.

We were therefore motivated to find an approach that would allow us to avoid the issues above mentioned. In what follows we show that by expanding algebraically in $\epsilon$ one can propose an approach that eliminates the explicit numerical dependance on $\epsilon$ while at the same time reducing the dimensionality of the Hilbert space needed for computation. 

\section{Frequency-filtered spectrum and photon correlations at zero delay time\label{sec:correlation at zero time delay}}

\subsection{$M=1$: power spectrum}

We begin by demonstrating the basics of our derivation by considering the emitter system coupled to only one sensor. Let us denote $\hat{\rho}_{ss}$ the steady state of the joint emitter-plus-sensor system. From \refeq{eq:n_i} we can calculate the power spectrum as:

\begin{equation}\label{eq:ps}
	S_{\Gamma_1}^{(1)}(\omega_1) = \frac{\Gamma_1}{2\pi \epsilon^2} \, \ev{n_1} = 
	\frac{\Gamma_1}{2\pi \epsilon^2} \, \textrm{Tr} \, [n_1\hat{ \rho}_{ss}].
\end{equation}
Considering the identity operator in the sensor Hilbert space i.e. $\mathbb{1}_{s_1}=\sum_{j_1=0,1}|j_1\rangle\langle j_1|$, we can write the full steady state $\hat{\rho}_{ss}$ as 
\begin{align}
\hat{\rho}_{ss} = \mathbb{1}_{s_1} \hat{\rho}_{ss} \mathbb{1}_{s_1} = \sum_{j_1,j_1'=0,1}  \hat{\rho}_{j_1,}^{j_1'} \otimes \ket{j_1} \bra{j_1'} \;,
\label{eq:SS_def_1sens}
\end{align}
where the matrices $\hat{\rho}_{j_1}^{j_1'}=\bra{j_1} \hat{\rho}_{ss} \ket{j'_1}$ are therefore only related to the degrees of freedom of the quantum emitter. Hermitian conjugates are obtained by swapping the upper and lower indices. Notice that each matrix  $\hat{\rho}_{j_1}^{j_1'}$ is thus of order $\epsilon^{j_1+j_1'}$. 
With this definition the power spectrum given in \refeq{eq:ps} becomes
\begin{equation}\label{eq:ps2}
	S_{\Gamma_1}^{(1)}(\omega_1) = \frac{\Gamma_1}{2\pi\epsilon^2} \, \mathrm{Tr} \,[\hat{\rho}_{1}^{1} ]
\end{equation}
To find the steady state, and in particular the matrix $\hat{\rho}_1^1$, we solve for the combined quantum emitter plus sensor system $\mathcal{L} (\hat{\rho}_{ss}) = 0$. We consider the action of the Liouvillian given in \refeq{eq:L} ($M=1$) on every term in \refeq{eq:SS_def_1sens}:
\begin{align}
\mathcal{L}(\hat{\rho}_{0}^{0} \otimes \ket{0} \bra{0}) = &\mathcal{L}_0(\hat{\rho}_{0}^{0})\otimes \ket{0} \bra{0} \label{eq:Lrho00}\\ 
&\nonumber -i\epsilon( a_1 \hat{\rho}_{0}^{0} \otimes \ket{1} \bra{0} -  \hat{\rho}_{0}^{0} a_1^{\dagger} \otimes \ket{0} \bra{1}), \; \label{eq:Lrho10}\\
 \mathcal{L}(\hat{\rho}_{1}^{0} \otimes \ket{1} \bra{0}) = &(\mathcal{L}_0 -\Gamma_1/2-i\omega_1)(\hat{\rho}_{1}^{0}\otimes \ket{1}\bra{0}   \\ &-i\epsilon( \nonumber 
a_1^{\dagger} \hat{\rho}_{1}^{0} \otimes \ket{0} \bra{0} -  \hat{\rho}_{1}^{0} a_1^{\dagger} \otimes \ket{1} \bra{1}) \label{eq:Lrho11}, \\
\mathcal{L}(\hat{\rho}_{1}^{1} \otimes \ket{1} \bra{1}) = &(\mathcal{L}_0 -\Gamma_1)(\hat{\rho}_{1}^{1}
\otimes \ket{1} \bra{1}) \\ 
+ \Gamma_1\hat{\rho}_{1}^{1}\otimes\ket{0} \bra{0} & -i\epsilon(\nonumber a_1^{\dagger} \hat{\rho}_{1}^{1} \otimes \ket{0} \bra{1} -  \hat{\rho}_{1}^{1} a_1 \otimes \ket{1} \bra{0}) \;,
\end{align}
and the expression for $\mathcal{L}(\hat{\rho}_{0}^{1} \otimes \ket{0_1} \bra{1_1})$ is the complex conjugate of \refeq{eq:Lrho10}. We can rewrite the sum of these expressions, in a similar way to \refeq{eq:SS_def_1sens}, by grouping together terms related to populations or coherences of the sensor:
\begin{subequations}
\begin{align}
\mathcal{L}(\hat{\rho}_{ss}) &= \sum_{j_1,j_1'=0,1}  \hat{B}_{j_1,}^{j_1'} \otimes \ket{j_1} \bra{j_1'}=\mathbf{0}\;, \\
\hat{B}_{j_1,}^{j_1'} &= \mathbf{0}, \; \mbox{For all }j_1,j_1'\;,
\label{eq:SS_zero}
\end{align}
\end{subequations}
In this way we can see our problem reduces to solving the set of coupled equations for $\hat{\rho}_{j_1,}^{j_1'}$ such that the operators $\hat{B}_{j_1,}^{j_1'}$ of the system are null matrices (zero at every element). Notice that \refeq{eq:Lrho00} has one term contributing to $\hat{B}_{0}^{0}$, which is zeroth order in $\epsilon$, and terms contributing to $\hat{B}_{1}^{0}$ and $\hat{B}_{0}^{1}$, which are of linear order in $\epsilon$. Similarly, \refeq{eq:Lrho11} contributes terms to $\hat{B}_{0}^{0}$, $\hat{B}_{1}^{1}$ as well as to $\hat{B}_{1}^{0}$ and $\hat{B}_{0}^{1}$ via a term proportional to $\epsilon$. Hence, the equation to be solve for $\hat{B}_0^0$, for instance, becomes
\begin{equation}
\hat{B}_0^0=\mathcal{L}_0(\hat{\rho}_{0}^{0}) - i\epsilon (a^{\dag}_1\hat{\rho}_1^0- \hat{\rho}_0^1 a_1) + \Gamma_1 \rho_1^1=\mathbf{0}.
\label{eq:B00}
\end{equation}
For an arbitrary value of $\epsilon$, the set of coupled equations given by \refeq{eq:SS_zero} does not have a simple solution. However, in the limit of weak coupling where $\epsilon \ll 1$ and $\ev{n_1}=\textrm{Tr}[\hat {\rho}_1^1]\ll 1$, we can neglect terms of the order of $\epsilon^2$. For instance, for $\hat{B}_0^0$ (\refeq{eq:B00} ) the terms $\Gamma_1\hat{\rho}_{1}^{1}$  and $\| i\epsilon(a_1^{\dagger} \hat{\rho}_{1}^{0}-\hat{\rho}_0^1 a_1)\|$ are of the order of $\epsilon^2$, so will be neglected. Likewise, in this weak coupling limit we have  $\|a_1^{\dagger} \hat{\rho}_{1}^{1}\| \ll \|\hat{\rho}_{0}^{0} a_1^{\dagger} \|$ in the equation for $\hat{B}_0^1$. These approximations can be generalised to a concept of ignoring down coupling, that is, the prefactor matrix $\hat{B}_{j_1,}^{j_1'}$ for $\ket{j_1} \bra{j'_1}$ has only contributions from terms $\hat{\rho}_{\ell}^{\ell'}$ with $\ell+\ell' \le j_1+j_1'$. This is equivalent to a formal expansion in $\epsilon$ as all the $\hat{\rho}_{\ell}^{\ell'}$ matrices  are of order $\epsilon^{\ell+\ell'}$. Using these approximations, we can write out the equations governing the steady state $\hat{B}_{j_1}^{j_1'}=0$ as
\begin{subequations}\label{eqs:SS_one_sensor}
\begin{align}
& \mathcal{L}_0 (\hat{\rho}_{0}^{0}) \sim 0 \\
&\mathcal{L}_0(\hat{\rho}_{1}^{0}) - (\Gamma_1/2 + i\omega_1)\hat{\rho}_{1}^{0}-i\epsilon a_1 \hat{\rho}_{0}^{0} \sim 0 \\
&\mathcal{L}_0(\hat{\rho}_{0}^{1}) - (\Gamma_1/2 -i\omega_1)\hat{\rho}_{0}^{1} +i\epsilon  \hat{\rho}_{0}^{0} a_1^{\dagger} \sim 0 \\
&\mathcal{L}_0(\hat{\rho}_{1}^{1}) - \Gamma_1 \hat{\rho}_{1}^{1} -i\epsilon  (a_1\hat{\rho}_{0}^{1}- \hat{\rho}_{1}^{0}a_1^{\dagger}) = 0 \; .
\end{align}
\end{subequations}
We can solve these equations in a chain from top to bottom, starting with $\hat{\rho}_{0}^{0}$. In practice, we need not solve for $\hat{\rho}_{0}^{1}$ as it is equal to $\hat{\rho}_{1 \phantom{\dagger}}^{0 \;\dagger}$.  Numerically we formulate the problem in Liouville space, such that $\lket{\hat{\rho}_{0}^{0}}$ is the zero eigenvector of the (square) matrix $\mathcal{L}_0$ given in \refeq{eq:L0}. The remaining equations can be solved as
\begin{subequations}\label{eqs:SS_1}
\begin{align}
\lket{\hat{\rho}_{1}^{0}} & \sim \frac{i\epsilon a_1 \lket{\hat{\rho}_{0}^{0} } }{\mathcal{L}_0 - (\Gamma_{1}/2 + i\omega_1)\mathbb{1}}  \\
\lket{\hat{\rho}_{1}^{1}}& = \frac{i\epsilon \left (a_1\lket{\hat{\rho}_{0}^{1}} -\lket{\hat{\rho}_{1}^{0}}{a_1}^{\dagger}\right )}{\mathcal{L}_0 - \Gamma_1 \mathbb{1}}   \; ,
\end{align}
\end{subequations}
$a_1$ and $a_1^\dagger$ are written in the Liouville space form, and ${\mathbb{1}}$ is the identity operator in the emitter Hilbert space. Notice \refeq{eqs:SS_1}(b) has an equality as for that case no term is discarded. In the above equations  $\hat{\rho}_{1}^{0}$ has a prefactor of $\epsilon$ and $\hat{\rho}_{1}^{1}$ has a prefactor of $\epsilon^2$. 
Therefore the dependence of the power spectrum (\refeq{eq:ps2}) on $\epsilon$ vanishes algebraically. The numerical calculation of the matrices given by Eqs. (\ref{eqs:SS_1}) can, in principle, be done by carrying over a small value for $\epsilon$. This procedure, however, could lead to numerical instabilities due to the smallness of $\epsilon$. 
With our method, such instabilities are prevented by computing the re-scaled matrices $\tilde{\hat{\rho}}_{j_1}^{j_1'} =  \hat{\rho}_{j_1}^{j_1'}/\epsilon^{j_1+j_1'}$ (such that $\tilde{\hat{\rho}}_{1}^{0} =  \hat{\rho}_{1}^{0}/\epsilon$ and $\tilde{\hat{\rho}}_{1}^{1} =  \hat{\rho}_{1}^{1}/\epsilon^2$), which are now $\epsilon$-independent system  operators. From the trace of the  $\epsilon$-independent matrix $\tilde{\hat{\rho}}_{1}^{1}$ we can calculate the sensor count rate as:
\begin{equation}
S_{\Gamma_1}(\omega_1) =  \frac{\Gamma_1}{2\pi} \mathrm{Tr}[\tilde{\hat{\rho}}_{1}^{1}]\,.
\end{equation}

\subsection{$M \geq 2$ zero-delay correlations}
The normalised second-order ($M=2$) photon correlation at zero delay time can be written as 
\begin{equation}\label{eq:g20}
	g_{\Gamma_1 \Gamma_2}^{(2)}(\omega_1,\omega_2) = 
	\frac{S_{\Gamma_1,\Gamma_2}^{(2)}(\omega_1,\omega_2)}{S_{\Gamma_1}^{(1)}(\omega_1) \, S_{\Gamma_2}^{(1)}(\omega_2)}
\end{equation}
where $S_{\Gamma_1}(\omega_1)$ and $S_{\Gamma_2}(\omega_2)$ are the mean count rates for the two sensors, as given in \refeq{eq:ps}, and:
\begin{equation}\label{eq:S2}
	S_{\Gamma_1,\Gamma_2}^{(2)}(\omega_1,\omega_2) = \frac{\Gamma_1 \Gamma_2}{(2\pi)^2 \epsilon^4} \,\ev{:n_1 n_2:}
\end{equation}
Since time-independent sensor number operators $n_j$ commute, normal order in \refeq{eq:S2} is unnecessary. Following the same procedure as before, we write our steady state density matrix, with two sensors included, as:
\begin{align}
\hat{\rho}_{ss} = \sum_{j_1,j_2,j_1',j_2'=0,1}  \hat{\rho}_{j_1,j_2}^{j_1',j_2'} \otimes \ket{j_1} \bra{j_1'} \otimes \ket{j_2} \bra{j_2'} \;
\label{eq:SS_def_two}
\end{align}
where $\{j_1, j_1'\}$ and $\{j_2, j_2'\}$ are counters over the states of sensor 1 and  sensor 2, respectively. As before, the matrices  
$\hat{\rho}_{j_1,j_2}^{j_1',j_2'}~=~\bra{j_1 j_2} \hat{\rho}_{ss} \ket{j'_1 j_2'}$ 
are defined in the Hilbert space of the quantum emitter alone and can be re-scaled as $\tilde{\hat{\rho}}_{j_1,j_2}^{j_1',j_2'} =\hat{\rho}_{j_1,j_2}^{j_1',j_2'}/\epsilon^{j_1+j_1'+j_2+j_2'}$. With this definition the second-order photon coincidence becomes
\begin{equation}\label{eq:S2_pert}
S_{\Gamma_1,\Gamma_2}^{(2)}(\omega_1,\omega_2) = \frac{\Gamma_1 \Gamma_2}{(2\pi)^2} \,\mathrm{Tr}\left [\tilde{\hat{\rho}}_{1,1}^{1,1} \right ], 
\end{equation}
with the power spectrum re-defined as
\begin{equation}\label{eq:S1}
	S_{\Gamma_1}^{(1)}(\omega_1) = \frac{\Gamma_1}{2\pi} \, \mathrm{Tr}[ \tilde{\hat{\rho}}_{1,0}^{1,0} ] \quad , \quad
	S_{\Gamma_2}^{(1)}(\omega_2) = \frac{\Gamma_2}{2\pi} \, \mathrm{Tr}[ \tilde{\hat{\rho}}_{0,1}^{0,1} ].
\end{equation}
To compute the matrices $\hat{\rho}_{j_1,j_2}^{j_1',j_2'}$ we solve for the steady state $\mathcal{L} (\hat{\rho}_{ss}) = 0$ with two sensors and by ignoring down coupling terms, that is, the matrix prefactor for $\ket{j_1} \bra{j_1'} \otimes \ket{j_2} \bra{j_2'}$ has only contributions from terms $\hat{\rho}_{\ell_1, \ell_2}^{\ell'_1, \ell'_2}$ satisfying the condition $\ell_1+\ell_2+\ell'_1+\ell'_2 \le j_1+j_2+j'_1+j'_2$. The resultant full set of linearly independent equations (besides those which are Hermitian conjugates of others) are then given by
\begin{subequations}
\label{eq:ss_derv}
\begin{align}
&\mathcal{L}_0 (\tilde{\hat{\rho}}_{0,0}^{0,0})\sim 0   \\
&[\mathcal{L}_0-\Gamma_1/2-i\omega_1] (\tilde{\hat{\rho}}_{1,0}^{0,0}) \sim i a_1 \tilde{\hat{\rho}}_{0,0}^{0,0}  \\
&[\mathcal{L}_0-\Gamma_2/2-i\omega_2] (\tilde{\hat{\rho}}_{0,1}^{0,0}) \sim i a_2 \tilde{\hat{\rho}}_{0,0}^{0,0}  \\
&[\mathcal{L}_0-\Gamma_1] (\tilde{\hat{\rho}}_{1,0}^{1,0}) \sim i (a_1 \tilde{\hat{\rho}}_{0,0}^{1,0}-  \tilde{\hat{\rho}}_{1,0}^{0,0}a_1^{\dagger}) \\
&[\mathcal{L}_0-\Gamma_2] (\tilde{\hat{\rho}}_{0,1}^{0,1}) \sim i (a_2 \tilde{\hat{\rho}}_{0,0}^{0,1}-  \tilde{\hat{\rho}}_{0,1}^{0,0}a_2^{\dagger}) \\ 
&[\mathcal{L}_0-(\Gamma_2+\Gamma_1)/2-i(\omega_1+\omega_2)] (\tilde{\hat{\rho}}_{1,1}^{0,0}) \sim \nonumber \\
&i (a_1 \tilde{\hat{\rho}}_{0,1}^{0,0}+ a_2 \tilde{\hat{\rho}}_{1,0}^{0,0}) \\
&[\mathcal{L}_0-(\Gamma_2+\Gamma_1)/2-i(\omega_1-\omega_2)] (\tilde{\hat{\rho}}_{1,0}^{0,1}) \sim \nonumber \\
&i (a_1 \tilde{\hat{\rho}}_{0,0}^{0,1}-  \tilde{\hat{\rho}}_{1,0}^{0,0}a_2^{\dagger}) \\
&[\mathcal{L}_0-(\Gamma_1/2+\Gamma_2)-i\omega_1] (\tilde{\hat{\rho}}_{1,1}^{0,1}) \sim \nonumber \\
&i (a_1 \tilde{\hat{\rho}}_{0,1}^{0,1} + a_2 \tilde{\hat{\rho}}_{1,0}^{0,1} - \tilde{\hat{\rho}}_{1,1}^{0,0}a_2^{\dagger}) \\
&[\mathcal{L}_0-(\Gamma_2/2+\Gamma_1)-i\omega_2] (\tilde{\hat{\rho}}_{1,1}^{1,0}) \sim \nonumber \\
&i (a_1 \tilde{\hat{\rho}}_{0,1}^{1,0}-\tilde{\hat{\rho}}_{1,1}^{0,0}a_1^{\dagger} + a_2 \tilde{\hat{\rho}}_{1,0}^{1,0} ) \\
&[\mathcal{L}_0-(\Gamma_1+\Gamma_2)] (\tilde{\rho}_{1,1}^{1,1}) = \nonumber \\
&i (a_1 \tilde{\hat{\rho}}_{0,1}^{1,1}-\tilde{\hat{\rho}}_{1,1}^{0,1}a_1^{\dagger} + a_2 \tilde{\hat{\rho}}_{1,0}^{1,1}-\tilde{\hat{\rho}}_{1,1}^{1,0}a_2^{\dagger} ) \; .
\end{align}\label{eqs:SS_two_sensor}
\end{subequations}
The solutions, in analogy to those in \refeq{eqs:SS_1}, are the following:
\begin{subequations}
\begin{align}
&\lket{\tilde{\hat{\rho}}_{1,0}^{0,0}} \sim \frac{i a_1 \lket{\tilde{\hat{\rho}}_{0,0}^{0,0} }}{\mathcal{L}_0 - (i\omega_1+\Gamma_{1}/2)\mathbb{1}},  \\ 
&\lket{\tilde{\hat{\rho}}_{0,1}^{0,0}} \sim \frac{i a_2 \lket{\tilde{\hat{\rho}}_{0,0}^{0,0} }}{\mathcal{L}_0 - (i\omega_2+\Gamma_{2}/2)\mathbb{1}}, \\
&\lket{\tilde{\hat{\rho}}_{1,0}^{1,0}} \sim \frac{i\, (a_1 \lket{\tilde{\hat{\rho}}_{0,0}^{1,0}} - \lket{\tilde{\hat{\rho}}_{1,0}^{0,0}} a_1^{\dag})}{\mathcal{L}_0 - \Gamma_{1}\mathbb{1}}, \\
&\lket{\tilde{\hat{\rho}}_{0,1}^{0,1}} \sim \frac{i\, (a_2 \lket{\tilde{\hat{\rho}}_{0,0}^{0,1}} - \lket{\tilde{\hat{\rho}}_{0,1}^{0,0}} a_2^{\dag})}{\mathcal{L}_0 -\Gamma_{2}\mathbb{1}}, \\
&\lket{\tilde{\hat{\rho}}_{1,1}^{0,0}} \sim \frac{i\, (a_1 \lket{\tilde{\hat{\rho}}_{0,1}^{0,0}} - a_2\lket{\tilde{\hat{\rho}}_{1,0}^{0,0}})}{\mathcal{L}_0 - (i\omega_1+i\omega_2+\Gamma_{1}/2+\Gamma_2/2)\mathbb{1}} , \\
& \lket{\tilde{\hat{\rho}}_{1,0}^{0,1}} \sim \frac{i\, (a_1 \lket{\tilde{\hat{\rho}}_{0,0}^{0,1}} - \lket{\tilde{\hat{\rho}}_{1,0}^{0,0}}a_2^{\dag})}{\mathcal{L}_0 - (i\omega_1-i\omega_2 +\Gamma_{1}/2+\Gamma_2/2)\mathbb{1}} , \\
& \lket{\tilde{\hat{\rho}}_{1,1}^{0,1}} \sim \frac{i\, (a_1 \lket{\tilde{\hat{\rho}}_{0,1}^{0,1}} + a_2\lket{\tilde{\hat{\rho}}_{1,0}^{0,1}} -\lket{\tilde{\hat{\rho}}_{1,1}^{0,0}} a_2^{\dag})}{\mathcal{L}_0 - (i \omega_1 +\Gamma_{1}/2+\Gamma_2)\mathbb{1}},\\
& \lket{\tilde{\hat{\rho}}_{1,1}^{1,0}} \sim \frac{i\, (a_1 \lket{\tilde{\hat{\rho}}_{0,1}^{1,0}} - \lket{\tilde{\hat{\rho}}_{1,1}^{0,0}}a_1^{\dag}+a_2\lket{\tilde{\hat{\rho}}_{1,0}^{1,0}} )}{\mathcal{L}_0 - (i\omega_2+\Gamma_{2}/2+\Gamma_1)\mathbb{1}},\\
& \lket{\tilde{\hat{\rho}}_{1,1}^{1,1}} =\frac{i\, (a_1 \lket{\tilde{\hat{\rho}}_{0,1}^{1,1}}- \lket{\tilde{\hat{\rho}}_{1,1}^{0,1}}a_1^{\dag}+a_2\lket{\tilde{\hat{\rho}}_{1,0}^{1,1}} - \lket{\tilde{\hat{\rho}}_{1,1}^{1,0}}a_2^{\dag})}{\mathcal{L}_0 - (\Gamma_1+\Gamma_2)\mathbb{1}} \, .
\end{align}
\end{subequations}

Generalisation of this formalism for the $M$-order frequency-resolved correlation at $\tau=0$ is straightforward and requires writing out the general steady state for the emitter and $M$ sensors in the form analogous to Eqs. \eqref{eq:SS_def_1sens} and \eqref{eq:SS_def_two}: 
\begin{equation}
\hat{\rho}_{ss} = \sum_{j_1,j_1',\ldots , j_M,j_M'= 0,1}  \hat{\rho}_{j_1\ldots j_m\ldots j_M}^{j'_1\ldots j_m'\dots j_M'}  \otimes \ket{j_1} \bra{j_1'} \otimes \ldots \otimes \ket{j_M}\bra{j_M'}  \; .
\label{eq:SS_def_ex}
\end{equation}
where $\{j_m, j_m'\}$ are counters over the state of  sensor $m$ and 
$\hat{\rho}_{j_1\ldots j_m \ldots j_M}^{j'_1\ldots j_m'\ldots j_M'} =\langle j_1 \ldots j_m\ldots j_M| \hat{\rho_{ss}} |j_1\ldots j_m'\ldots j_M' \rangle$. 
We define the re-scaled matrices
$\tilde{\hat{\rho}}_{j_1\ldots j_m \ldots j_M}^{j'_1\ldots j_m'\ldots j_M'}=\hat{\rho}_{j_1\ldots j_m \ldots j_M}^{j'_1\ldots j_m'\ldots j_M'}/ \epsilon^{j_1+j_1'+\ldots+j_m+j_m'\dots + j_M+j_M}$.
The $M$th order photon-coincidence at zero-delay time is given in terms of  the trace of  matrix with $j_m=j_m'=1$ for all $m$
\begin{align}
S_{\Gamma_1 \ldts \Gamma_M}^{(M)}(\omega_1,... \omega_m,...\omega_M)=\frac{\Gamma_1...\Gamma_m...\Gamma_M}{(2\pi)^M} \mathrm{Tr}\left [\tilde{\hat{\rho}}_{1\ldots1\ldots 1}^{1\ldots1\ldots 1}\right ],
\label{eq:SMgeneral}
\end{align}
and the power spectrum for each sensor $m$ is given by the trace of the matrix with $j_m=j_m'=1$ for $m$ and $j_y=j_y'=0$ from $y\neq m$:
\begin{align}
S_{\Gamma_m}^{(1)}(\omega_m) = \frac{\Gamma_m}{2\pi} \, \mathrm{Tr}\left [ \tilde{\hat{\rho}}_{0\dots 1\dots 0}^{0\ldots 1\ldots 0} \right ].
\label{eq:spectrumgeneral}
\end{align} 
The general equation satisfied by the matrices $\tilde{\hat{\rho}}_{j_1 \ldots j_m\dots j_M}^{j_1' \ldots j_m'\dots j_M'}$ such that $\mathcal{L} (\hat{\rho}_{ss}) = 0$ becomes:
\begin{align}
\label{eq:general}
&\left[\mathcal{L}_0-\sum_{m=1}^M \left \{(j_m+j'_m)\Gamma_m/2+(j_m-j'_m)i\omega_m)\right \} \right] \tilde{\hat{\rho}}_{j_1 \ldots j_m\dots j_M}^{j_1' \ldots j_m'\dots j_M'} \nonumber \\
&= i \sum_{m=1}^M \left [ \delta_{j_m,1}{a}_m \tilde{\hat{\rho}}_{j_1 \ldots j_m(1 -\delta_{j_m,1})\ldots j_M}^{j_1' \ldots j_m' \ldots j_M'}
- \delta_{j_m',1} \tilde{\hat{\rho}}_{j_1 \ldots j_m\ldots j_M}^{j_1' \ldots j_m' (1-\delta_{j_m',1}) \ldots j_M'} a_m^{\dagger}\right ].
\end{align}
Here $\delta_{u,v}$ is the Kronecker delta function, equal to zero if $u \neq v$ or unity if $u=v$. The derivation of \refeq{eq:general} is discussed in Appendix \ref{app:M}.

We conclude this section by highlighting that our approach to compute multi-photon correlations in the frequency domain is quite efficient as it depends on the Liouvillian of the emitter alone $\mathcal{L}_0$. This should provide an important advantage for quantum emitters of large Hilbert space dimension. Notice also that, while we have assumed a Lindblad form for $\mathcal{L}_0$  (See \refeq{eq:Lm}), the relation in \refeq{eq:general} does not explicitly depends on this fact. Hence, if one is able to generalise the proof for the  equivalence between the  sensor method and the integral methods beyond the Markovian and quantum regression restriction presented in the supplemental material of \cite{DelValle2012}, our result in \refeq{eq:general} will apply to open quantum systems undergoing non-Markovian, non-pertubative dynamics.

\section{Frequency-filtered correlations at finite delay time\label{sec:tdep_pert}}

In this section we will use time-dependent perturbation theory to construct solutions for the correlation functions at finite time delay. We focus on the second-order correlation function at finite delay denoted $g_{\Gamma_1 \Gamma_2}^{(2)}(\omega_1,T,\omega_2,T+\tau)$. In the steady state $\hat{\rho}_{ss}$, the explicit time dependence on $T$ vanishes and we can simply write $g_{\Gamma_1 \Gamma_2}^{(2)}(\omega_1, \omega_2, \tau)$. In terms of the sensor operators, this correlation is expressed as 
\begin{equation}
g_{\Gamma_1 \Gamma_2}^{(2)}(\omega_1, \omega_2, \tau) = \frac{S_{\Gamma_1,\Gamma_2}^{(2)}(\omega_1,\omega_2,\tau)}{S_{\Gamma_1}^{(1)}(\omega_1) \, S_{\Gamma_2}^{(1)}(\omega_2)}, 
\label{eq:g2_tau}
\end{equation}
where the numerator is given by
\begin{equation}\label{eq:S2tau}
S_{\Gamma_1,\Gamma_2}^{(2)}(\omega_1,\omega_2,\tau) = \frac{\Gamma_1 \Gamma_2}{(2\pi)^2 \epsilon^4} \; \ev{\varsigma_1^{\dagger} \varsigma_2^{\dagger}(\tau) \varsigma_2(\tau),\varsigma_1}
\end{equation}
and the functions in the denominator are given in \refeq{eq:S1}. We consider the correlations in  \refeq{eq:S2tau} when $\tau>0$, meaning sensor 1 first registers a detection, then sensor 2 does so a time $\tau$ later. Here the normal time ordering is of crucial importance as the sensor operators do not commute at different times. Correlations for $\tau <0$ are obtained by exchanging $\varsigma_1\rightarrow \varsigma_2$ and  $\varsigma_2(\tau)\rightarrow\varsigma_1(\tau)$. Expressed in Liouville space, the correlation given in \refeq{eq:S2tau} is written as 
\begin{align}\begin{split}
\ev{\varsigma_1^{\dagger} \varsigma_2^{\dagger}(\tau) \varsigma_2(\tau) \varsigma_1} \,
&= \text{Tr}\{\varsigma_2^{\dag}\varsigma_2 \, \mathcal{G}(\tau) \varsigma_1 \hat{\rho}_{ss} \varsigma_1^{\dagger}\} \\
&\equiv \lbra{\varsigma_2^{\dagger}\varsigma_2} \mathcal{G}(\tau) \varsigma_1 \hat{\rho}_{ss} \varsigma_1^{\dagger}\drangle \;,
\label{eq:g2tau}
\end{split}\end{align}
with $\mathcal{G}(\tau)=\exp(\mathcal{L}t)$, the time propagator operator for the joint sensor plus emitter with $\mathcal{L}$ given in \refeq{eq:L}. 
The term $\varsigma_1 \hat{\rho}_{ss} \varsigma_1^{\dagger}$ represents a photon detection on sensor 1 which resets this sensor to its ground state and leaves the emitter and sensor 2 in a joint ``conditional state", that is, 
\begin{align}
\hat{\rho}(0) = \varsigma_1 \hat{\rho}_{ss} \varsigma_1^{\dagger} =  \sum_{j_2,j_2'=0,1}  \hat{\rho}_{1,j_2}^{1,j_2'} \otimes \ket{j_2} \bra{j_2'} \otimes \ket{0_1} \bra{0_1}   \;.
\label{eq:initcond}
\end{align}
Notice that $\hat{\rho}(0)$ is not normalised but has a trace equal $\ev{\varsigma_1^{\dagger} \varsigma_1}$. Let us define $\lket{\hat{\rho}(\tau)}= \mathcal{G}(\tau) \lket{\hat{\rho}(0)}$ with the initial condition $\lket{\hat{\rho}(0)} = \lket{\varsigma_1\hat{ \rho}_{ss} \varsigma_1^{\dagger}}$. In principle, one can perform this explicit time-propagation. Since sensor 1 is now in the ground state, only the interaction Hamiltonian with sensor 2, $H_{e,2} = \epsilon\, (a_2 \varsigma_2^{\dagger} +a_2^{\dagger}\varsigma_2)$, contributes to the joint dynamics and the propagation requires to test for convergence in $\epsilon$. Alternatively, since we are interested in the regime where $\epsilon$ is small, we can evaluate such dynamics by using time-dependent perturbation theory with respect to $H_{e,2}$. We then proceed to expand $\lket{\hat{\rho}(\tau)}$ as \cite{MukamelBook}
\begin{equation}\label{rho_pert}
	\lket{\hat{\rho}(\tau)} = \lket{\hat{\rho}^{(0)}(\tau)}+\lket{\hat{\rho}^{(1)}(\tau)}+\lket{\hat{\rho}^{(2)}(\tau)}+\ldots
\end{equation}
The zeroth order term corresponds to the dynamics given by the emitter and sensors Liouvillians without interaction, that is, 
 $\lket{\hat{\rho}^{(0)}(\tau)~}=~\mathcal{G}_0(t)\lket{\hat{\rho}(0)}$
with $\mathcal{G}_0(t) = \exp([\mathcal{L}_0+\mathcal{L}_1+\mathcal{L}_2] t)$
and $\mathcal{L}_0, \mathcal{L}_1$ and $\mathcal{L}_2$ as in Eqs. \eqref{eq:L0} and \eqref{eq:Lm}. For the initial condition considered, the sensor 1 (2) will not contribute to the dynamics for $\tau>0$ ($\tau<0$) and can thus be traced over. The $k$th order solution requires $k$ interactions with $H_{e,2}$, but time propagation occurs only in terms of $\mathcal{G}_0(t)$: 
\begin{equation}
\lket{\hat{\rho}^{(k)}(\tau)}=-i \int_0^{\tau} dt\,H_{e,2}^{\times} \mathcal{G}_0(\tau-t)\lket{\hat{\rho}^{(k-1)}(t)},
\end{equation}
with $H_{e,2}^{\times}\lket{\rho} \equiv [H_{e,2},\rho]$ denoting a commutator superoperator in Liouiville space. Notice that  $\hat{\rho}(0)$ contains  $\hat{\rho}_{1,j_2}^{1,j_2'}$ which has terms of order $\epsilon^{2+j_2+j_2'}$, that is, from $\epsilon^2$ up to $\epsilon^4$. This indicates that to compute the second order correlation function we need to consider up to second order perturbation theory as this will be the first term that, by requiring two iterations of $H_{e,2}$, will be of the same order $\epsilon^2$. Third order perturbation will result in terms of the order of $\epsilon^5$ or higher, which are negligible under the weak coupling assumption. The time-resolved photon correlation can thus be written as
\begin{align}
\ev{\varsigma_1^{\dagger} \varsigma_2^{\dagger}(\tau) \varsigma_2(\tau) \varsigma_1} = 
\langle\mkern-3mu\langle\varsigma_2^{\dagger}\varsigma_2 \lket{\hat{\rho}(\tau)} = I_0(\tau) + I_1(\tau) + I_2(\tau)
\label{eq:g2taubis}
\end{align}
where $I_k(\tau)=\langle \mkern-3mu \langle\varsigma_2^{\dagger}\varsigma_2 \lket{\hat{\rho}^{(k)}(\tau)}=\mathrm{Tr}\{\varsigma_2^{\dagger}\varsigma_2 \hat{\rho}^{(k)}(\tau)\}$.

The \emph{zeroth order} term becomes
\begin{align}
I_0(\tau) = \lbra{n_2} \mathcal{G}_0(\tau) {\hat{\rho} (0)} \drangle =
e^{-\Gamma_2 \tau} \mathrm{Tr}\left [ \hat{\rho}_{1,1}^{1,1} \right ] \;.
\label{eq:zeroorder}
\end{align}
Here it is relevant to notice that this term contains the same information as $g^{(2)}(0)$ at zero time delay (see \refeq{eq:g20} and \refeq{eq:S2_pert}), while the exponential time-dependence provides no information about the emitter dynamics as it only relates to the uncertainty in detection time.\\
The next term arises from the \emph{first order} perturbation theory, in the form
\begin{align}
I_1(\tau) = -i \int_0^{\tau}  dt_1 \lbra{n_2} \mathcal{G}_0(\tau-t_1) H_{e,2}^{\times}\mathcal{G}_0(t_1) \hat{\rho}(0)\drangle  \;.
\end{align}
We can act the first $\mathcal{G}_0(\tau-t_1)$ to the left, giving $\lbra{n_2} \mathcal{G}_0(\tau-t_1) =  \lbra{n_2} \exp[-\Gamma_2(\tau-t_1) ] $. The only elements to the right which will contribute are $a_2 \hat{\rho}_{1,0}^{1,1} (t_1) \exp[-(\Gamma_2/2+i \omega_2) t_1] $ and $\hat{\rho}_{1,1}^{1,0}(t_1)a_2^{\dagger} \exp[-(\Gamma_2/2-i \omega_2) t_1]$, with $\lket{\hat{\rho}_{1,j_2}^{1,j_2'}(t)}~ \equiv~ \Theta[t]\exp(\mathcal{L}_0 t)\lket{\hat{\rho}_{1,j_2}^{1,j_2'}(0)}$ defined via the evolution of the emitter alone. As these two terms are complex conjugates, we can write: 
\begin{align}
I_1(\tau) =  2 \epsilon \; \mathrm{Im}\left(  \int_0^{\tau}  dt_1 e^{-\Gamma_2 (\tau-t_1/2)+i\omega_2 t_1}  \mathrm{Tr}[ a_2 \hat{\rho}_{1,0}^{1,1}(t_1)] \right)\;,
\label{eq:firstorder}
\end{align} 
This is essentially a finite time Laplace transform of a complex number, which is simple enough to perform numerically. Here the density matrix $\hat{\rho}_{1,0}^{1,1}(t_1)$ evolves under the action of Liouvillian $\mathcal{L}_0$.  

Finally, the \emph{second order} term, $I_2(\tau)$, reads
\begin{align}
I_2(\tau) = -\int_0^{\tau}  dt_2  &\int_0^{t_2}  dt_1 \; \lbra{n_2} \mathcal{G}_0(\tau-t_2)  \nonumber \\ 
&H_{e,2}^{\times}\mathcal{G}_0(t_2-t_1) H_{e,2}^{\times} \mathcal{G}_0(t_1) \lket{\hat{\rho}(0)} \;.
\label{eq:I2tau}
\end{align}
Because we have two applications of $H_{e,2}$,  both $\hat{\rho}_{1,0}^{1,0}\otimes \ket{0_2} \bra{0_2}$ and $\hat{\rho}_{1,1}^{1,1}\otimes \ket{1_2} \bra{1_2}$ terms can contribute. However, since $\mathrm{Tr}[\hat{\rho}_{1,1}^{1,1}] \ll \mathrm{Tr}[\hat{\rho}_{1,0}^{1,0}]  $, we need only to consider $\hat{\rho}_{1,0}^{1,0} \otimes \ket{0_2} \bra{0_2}$ in our initial condition. Hence, to lowest order in $\epsilon$, we have
\begin{align}
I_2(\tau) = -\epsilon^2 &\int_{0}^{\tau}  dt_2  \int_0^{t_2}  dt_1 e^{-\Gamma_2(\tau-t_2)} \nonumber \\
&\lbra{n_2} \overrightarrow{a_2\varsigma_2^{\dagger}} \mathcal{G}_0(t_2-t_1) \overleftarrow{a_2^{\dagger} \varsigma_2}  \mathcal{G}_0(t_1) \lket{\rho(0)} + h.c.\nonumber \\
= -2\epsilon^2 \mathrm{Re}\; &\int_{t_1}^{\tau}  dt_2  \int_{0}^{t_2}  dt_1 e^{-\Gamma_2[\tau-(t_2+t_1)/2]+i\omega_2(t_2-t_1)}  \nonumber \\
&\mathrm{Tr} \{ a_2(t_2-t_1)  \hat{\rho}_{1,0}^{1,0}(t_1) a_2^{\dagger} \} \;,
\label{eq:secondorder2}
\end{align}
with $h.c$ meaning Hermitian conjugate, $\overrightarrow{O} \lket{\rho}\equiv \hat{O} \hat{\rho}$ and $\overleftarrow{O} \lket{\rho}\equiv \hat{\rho} \hat{O} $. We have used the Heisenberg and Schr\"odinger pictures such that $a_2(t)$ is the time dependent operator, propagating under the adjoint of $\mathcal{L}_0$. The double integral over the two dimensional simplex is numerically more complex, but can be performed. 

\refeq{eq:S2tau} for the time-resolved two-photon coincidence becomes
\begin{equation}
S_{\Gamma_1,\Gamma_2}^{(2)}(\omega_1,\omega_2,\tau>0) = \frac{\Gamma_1\Gamma_2}{(2\pi)^2 \epsilon^4}\left[I_0(\tau) +I_1(\tau) + I_2(\tau)\right] \;.
\label{eq:s2_tau_SI}
\end{equation}
Notice that $I_0(\tau)$, $I_1(\tau)$ and $I_2(\tau)$ will all feature a prefactor $\epsilon^4$. Hence the $\epsilon$ dependence in \refeq{eq:s2_tau_SI} cancels out algebraically, as expected. The time-resolved photon-coincidence can therefore be written as
\begin{equation}
S_{\Gamma_1,\Gamma_2}^{(2)}(\omega_1,\omega_2,\tau>0) = \frac{\Gamma_1\Gamma_2}{(2\pi)^2 }\left[\tilde{I}_0(\tau) +\tilde{I}_1(\tau) + \tilde{I}_2(\tau)\right] \;,
\label{eq:s2_tau}
\end{equation}
with $\tilde{I}_k(\tau) = \epsilon^{-4} I_k(\tau)$ the $k$th order term, which requires $k$ interactions with the coupling Hamiltonian $H_{e,2}$.  
The final expression for the second-order correlation at a finite time delay reads
\begin{align}
g_{\Gamma_1 \Gamma_2}^{(2)}(\omega_1, \omega_2, \tau>0)= \frac{\tilde{I}_0(\tau) +\tilde{I}_1(\tau) + \tilde{I}_2(\tau)}{\ev{n_1}\ev{n_2}}
\label{eq:g2taufinal}
\end{align}
with $\ev{n_1}=\textrm{Tr}\left [ \tilde{\hat{\rho}}_{1,0}^{1,0}\right ]$  and $\ev{n_2}=\textrm{Tr}\left [ \tilde{\hat{\rho}}_{0,1}^{0,1}\right ]$. The correlation for $\tau<0$ is obtained by taking $\hat{\rho}(0) = \varsigma_2 \hat{\rho}_{ss} \varsigma_2^{\dagger}$ and doing time-dependent perturbation theory with respect to $H_{e,1}$. This results in making the replacements $\Gamma_2 \to \Gamma_1$, $\omega_2 \to \omega_1$,  $a_2\to a_1$, $\hat{\rho}_{1,0}^{1,1}(t_1)\to \hat{\rho}_{0,1}^{1,1}(t_1)$, and $\hat{\rho}_{1,0}^{1,0}(t_1)\to \hat{\rho}_{0,1}^{0,1}(t_1)$ in Eqs.\eqref{eq:zeroorder}, \eqref{eq:firstorder} and \eqref{eq:secondorder2}. At this point it is worth noting that,  in general, the time-resolved correlation function can exhibit time asymmetry whenever the two frequencies detected are different from each other ($\omega_1\neq\omega_2$), even if $a_1=a_2$. This is evident from the definition of $I_1$ in \refeq{eq:firstorder} and the definition of $I_2$ in \refeq{eq:secondorder2}: both $I_1$ and $I_2$ have exponentials in their integrands that explicitly depend on $\omega_2$ or $\omega_1$ for positive or negative times, respectively. Similar arguments apply to the specific matrix operators involved for the different time regimes. Time-symmetrical functions are expected when we have identical system emission operators ($a_1=a_2$), identical frequencies ($\omega_1=\omega_2$) and identical sensor decay rates ($\Gamma_1=\Gamma_2$).
\\
\subsection{Behaviours of at short- and large-time delays\label{sec:approximations}}
We first consider the short-time delay regime. As we discussed above, $\tilde{I}_0{\tau} \propto e^{-\Gamma_2 \tau}$ for all $\tau \ge 0$ indicating that its time dependence simply captures the uncertainty in the detection. When $\tau$ is smaller than any relevant system timescales, we have, to lowest order $\tilde{I}_1(\tau) \sim 2\tau \, \mathrm{Im}(\mathrm{Tr} [ a_2  \tilde{\hat{\rho}}_{1,0}^{1,1} ])$ and $\tilde{I}_2(\tau) ~\sim ~\tau^2 \, \mathrm{Re} \big (\mathrm{Tr} [ a_2  \tilde{\hat{\rho}}_{1,0}^{1,0} a_2^{\dagger} ] \big )$. The most interesting information is thus given by the short time behaviour of $\tilde{I_2}(\tau)$. Since this function involves a propagation in time after a first iteration with $H_{e,2}$ (see \refeq{eq:I2tau}), its short-time behaviour can have contributions from coherent dynamics within the excited manifold of the system of interest. In fact, the proportionality of $\tilde{I}_2(\tau)$ to $\tau^2$ is suggestive that quantum speed-up processes are being captured by this function \cite{Cimmarusti2015}. For $\tau < 0$, the sensor ordering is reversed and we have instead: $\tilde{I}_1(\tau) \sim 2\tau \, \mathrm{Im}(\mathrm{Tr} \{ a_1  \tilde{\hat{\rho}}_{0,1}^{1,1} \})$ and $\tilde{I}_2(\tau) \sim \tau^2 \, \mathrm{Re}(\mathrm{Tr} \{ a_1  \tilde{\hat{\rho}}_{0,1}^{0,1} a_1^{\dagger} \})$. In general, $\mathrm{Re}( \mathrm{Tr} \{ a_1  \tilde{\hat{\rho}}_{0,1}^{0,1} a_1^{\dagger} \}) \neq \mathrm{Re}(\mathrm{Tr} \{ a_2  \tilde{\hat{\rho}}_{1,0}^{1,0} a_2^{\dagger} \})$ and likewise  $\mathrm{Im}(\mathrm{Tr} \{ a_2  \tilde{\hat{\rho}}_{1,0}^{1,1} \}) \ne \mathrm{Im}(\mathrm{Tr} \{ a_1  \tilde{\hat{\rho}}_{0,1}^{1,1} \})$. Hence, we will expect an asymmetry in $g^{(2)}_{\Gamma_1, \Gamma_2}(\omega_1, \omega_2, \tau)$ for positive and negative $\tau$, even in the case when $a_1=a_2$.

We now investigate $\tilde{I}_1(\tau)$ and $\tilde{I}_2(\tau)$ in the regime where $\tau$ becomes large relative to the emitter or sensor linewidth timescales. Let us call $\gamma_{sys}$ the largest  emitter decay rate linked to the field operator $a_2$. If $\gamma_{sys} \gg \Gamma_2$ and $\tau \gamma_{sys} \gg 1$, we can make the approximation
\begin{align}
\tilde{I}_1(\tau) \sim  2 e^{-\Gamma_2 \tau} \; \mathrm{Im}\left(  \int_0^{\infty}  dt_1 e^{+\Gamma_2 t_1/2+i\omega_2 t_1}  \mathrm{Tr}[ a_2{\tilde{\hat{\rho}}_{1,0}^{1,1}}(t_1)] \right), 
\label{eq:firstorder_limit1}
\end{align}
where the integral, now independent of $\tau$, can be identified as the infinite Laplace transform $F(s)$ of $\mathrm{Tr}[ a_2\tilde{\rho}_{1,0}^{1,1}(t_1)]$ and $s~=~\Gamma_2/2+i\omega_2$, i.e. $\tilde{I}_1(\tau) \sim 2e^{-\Gamma_2 \tau} \mathrm{Im}\{F(\Gamma_2/2+i\omega_2)\}$; thus time dependence only due to uncertainty in the detection time. Since $I_1(0)=0$, we expect the full form of $\tilde{I}_2(\tau)$ to undergo an initial rise, followed by an exponential decay. 
On the other hand, if $\gamma_{sys} \ll \Gamma_2$, we can approximate $\tilde{\hat{\rho}}_{1,0}^{1,1}(t)$ as having a single dominant coherent transition frequency $\omega_{sys}$, that is, $\tilde{\hat{\rho}}_{1,0}^{1,1}(t)\simeq \exp(+\gamma_{sys}t - i\omega_{sys}t) \tilde{\hat{\rho}}_{1,0}^{1,1}(t)$ and slowly varying. Let us define $\tilde{t_1} = \tau - t_1$ so we can write
\begin{align}
\tilde{I}_1(\tau) =  &2 \; \mathrm{Im}\left( \int_0^{\tau}  d\tilde{t}_1 e^{-\Gamma_2 (\tilde{t}_1+\tau)/2+i\omega_2 (\tilde{t}_1-\tau)}  \mathrm{Tr}[ a_2 \tilde{\hat{\rho}}_{1,0}^{1,1}(\tau-\tilde{t}_1)] \right) \nonumber \\
\sim  &2 \; \mathrm{Im}\left(  \frac{e^{-(\Gamma_2+\gamma_{sys})\tau/2-i(\omega_{sys}-\omega_2) \tau}-e^{-\Gamma_2\tau}}{(\Gamma_2-\gamma_{sys})/2+i(\omega_2-\omega_{sys})} \mathrm{Tr}[ a_2\tilde{\hat{\rho}}_{1,0}^{1,1}(\tau)] \right)  \;,
\label{eq:firstorder_limit2} 
\end{align}
where, by assumption, the dominant term is the numerator of the fraction resulting in a damped oscillatory function. The approximation of a single frequency breaks down when the sensor linewidth is smaller than the emission spectrum.

We expect $g^{(2)}_{\Gamma_1, \Gamma_2}(\omega_1, \omega_2, \tau) \to 1$ when $\Gamma_2 \tau \gg $ and $\gamma_{sys}\tau \gg $. Since $\tilde{I}_0(\tau)$ and $\tilde{I}_1(\tau)$ decay exponentially in this regime, $\tilde{I}_2(\tau)$ must therefore tend to a constant value. To see this we rewrite $\tilde{I}_2$ in terms of $\tilde{t}_1=\tau-(t_2+t_1)/2$ and $\tilde{t}_2 = t_2 - t_1$ as
\begin{align}
\tilde{I}_2(\tau) = &2 \mathrm{Re} \int_0^{\tau} d\tilde{t}_2 \; \int_0^{\tau-\tilde{t}_2/2} d\tilde{t}_1 \; e^{-\Gamma_2\tilde{t}_1 + i\omega_2 \tilde{t}_2} \nonumber \\
&\mathrm{Tr} \{ a_2(\tilde{t}_2) \tilde{\hat{\rho}}_{1,0}^{1,0}(\tau - \tilde{t}_1 - \tilde{t}_2/2) a_2^{\dagger} \} \;.
\label{eq:secondorder3}
\end{align}
As $\tau \to \infty$, $\tilde{\hat{\rho}}_{1,0}^{1,0}(\tau)$ will approach the functional form of our original steady state for the emitter, and so we can write $\tilde{\hat{\rho}}_{1,0}^{1,0}(\tau - \tilde{t}_1/2 - \tilde{t}_2)  \to \ev{n_1}[\tilde{\hat{\rho}}_{0,0}^{0,0}-\Delta\tilde{\hat{\rho}}_{ss}(\tau - \tilde{t}_1/2 - \tilde{t}_2)]$. We also expect the trace of the difference term $\Delta\rho_{ss}$ to be exponentially small when $\tau \to \infty$, and the variation in terms of $\tilde{t}_1$ and $\tilde{t}_2$ to be slow enough to neglect. We can therefore take the integral over $\tilde{t}_1$ and obtain
\begin{align}
\tilde{I}_2(\tau) \sim  \frac{2 \ev{n_1}}{ \Gamma_2}  \mathrm{Re} & \int_0^{\tau}  d\tilde{t}_2 \; \left( 1- e^{-\Gamma_2(\tau -\tilde{t}_2/2)} \right) e^{+i\omega_2 \tilde{t}_2} \nonumber \\
&\mathrm{Tr} \left [ a_2(\tilde{t}_2)  \big \{ \tilde{\rho}_{0,0}^{0,0}-\Delta\tilde{\rho}_{ss}(\tau- \tilde{t}_2/2)\big \} a_2^{\dagger} \right ] \;.
\label{eq:secondorder4}
\end{align}
If we take the integral over $\tilde{t}_2$ to infinity (assuming $\gamma_{sys}\tau \gg 1$), the term dependent on $\tilde{\hat{\rho}}_{0,0}^{0,0}$ will tend to $\ev{n_1}\ev{n_2}$ and the remainder term, which is a function of $\Delta\tilde{\rho}_{ss}(\tau- \tilde{t}_2/2)$, tends to zero, giving $g^{(2)}_{\Gamma_1, \Gamma_2}(\omega_1, \omega_2, \tau) \to 1$. Assuming $\Delta\tilde{\rho}_{ss}(t)$ does not have rapidly oscillating components, we expect \refeq{eq:secondorder4} to be a good general approximation for a wide range of $\tau$.

We conclude this section by noting that the second-order time-dependent perturbation approach described above can also be used in the calculation of higher-order photon correlations when only one sensor has a time delayed detection, i.e. $g^{(M)}_{\Gamma_1,...\Gamma_m...\Gamma_M)}(\omega_1,...\omega_m, \tau....\omega_M)$. The generalisation to multiple time delays is, however, more elaborated. For instance, computation of the 3rd order correlations for different delay times will require the application of second-order perturbation theory with respect to the interactions with sensors 2 and 3,  $H_{e,2}$  and $H_{e,3}$ respectively, but at different stages in the dynamics. This will lead to a four dimensional numerical integration. In this case one can instead take advantage of the efficient method to compute the auxiliary matrices defining the steady state (see \refeq{eq:general}) and propagate in time without perturbation.

\section{Comparison of original and proposed method \label{sec:numerics}}

To compare our approach and the original formulation of the sensor method, we consider a toy model for a prototype vibronic dimer  as in Ref. \cite{OReilly2014, JCP2015_Killoran, Dean2016}. We are motivated by the experimental measurements of room-temperature photo-counting statistics of similar bichromophoric \cite{PNAS03_Hofkens, Hubner2003} and multi-chromophoric systems \cite{Wientjes2014} systems, all of which have given evidence of anti-bunching. In our toy model, each chromophore has an excited electronic state $|k\rangle$ with energy $\alpha_k$, $k=1,2$, and is locally coupled with strength $g$ to a quantized vibrational mode of frequency $\omega_\textrm{vib}$. Inter-chromophore coupling is generated by dipole-dipole interaction of strength $V$. The electronic Hamiltonian and the bare vibrational Hamiltonian read
$ H_{\textrm{el}} = \alpha_1 |1\rangle\langle 1|+ \alpha_2 |2\rangle\langle 2|+V(|1\rangle\langle 2|+|2\rangle\langle 1|)$,
and 
$H_{\textrm{vib}}= ~\omega_{\textrm{vib}}(d_1^\dag d_1+ ~d_2^\dag d_2)$,
respectively. Linear coupling of  electronic excited states to their corresponding local vibration is described via
$ H_{\textrm{el-vib}} ~=~ g \sum_{k=1,2}|k\rangle\langle k|(d^\dag_k+ d_k)$,
where $d^\dag_k$($d_k$) creates (annihilates) a phonon of the vibrational mode of chromophore $k$. We denote $|X_1\rangle$ and $|X_2\rangle$ the exciton eigenstates of  
$H_{\textrm{el}}$, with corresponding eigvalues $E_1$ and $E_2$ yielding an average energy $E=(E_1+E_2)/2$ and a splitting 
$\Delta E =\sqrt{(\Delta \alpha)^2+4V^2}$, with $ \Delta \alpha=\alpha_1-\alpha_2$
being the positive difference between the onsite energies. Including the electronic ground state $|G\rangle$ with energy set to zero, we define the electronic basis $\{|G\rangle, |X_1\rangle, |X_2\rangle\}$. Transformation of this electronic-vibrational Hamiltonian into normal mode coordinates \cite{MayKhun,OReilly2014} shows that only the 
relative displacement mode, with creation operator 
$D^{(\dag)}=(d^{(\dag)}_1- d^{(\dag)}_2)/\sqrt{2}$,
couples to the excitonic system. The effective Hamiltonian for the prototype dimer takes the form of a generalised quantum Rabi model \cite{Xie2017}:
\begin{equation}\begin{split}\label{H_diag}
	H_0 = & \, E \, \tilde{M} + \frac{\Delta E}{2} \: \tilde{\sigma}_z + 
	\omega_{vib} D^\dag D\\ 
	& \; + \frac{g}{\sqrt{2}}   \: 
	\Big(\cos(2\theta) \, \tilde{\sigma}_z - \sin(2\theta) \, \tilde{\sigma}_x\Big) (D+D^\dag).
\end{split}\end{equation}
Here we have defined the collective electronic operators
$\tilde{M}~=~ |X_1\rangle \langle X_1| + |X_2\rangle\langle X_2|$,
$\tilde{\sigma_z}= |X_1\rangle \langle X_1| - |X_2\rangle\langle X_2|$, 
and 
$\tilde{\sigma}_x =|X_2\rangle \langle X_1| +|X_1\rangle \langle X_2| $, 
and the mixing angle $\theta~=~1/2 \arctan \left(2|V|/\Delta\alpha \right)$ satisfies $0<\theta<\pi/4$.  The vibrational eigenstates of $D^\dag D$ are denoted as $|l\rangle$, which for the purpose of numerical computation, are set to a maximum number $L$, i.e. $l=0,1,\cdots,L$. Hence, the ground electronic-vibrational eigenstates of $H_0$ are of the form $|G, l\rangle\equiv|G\rangle\otimes |l\rangle$ while the excited vibronic eigenstates, labelled $|F_v\rangle$, can be written as quantum superpositions of states $|X_i,l\rangle\equiv|X_i\rangle\otimes| l\rangle$ i.e. $|F_v\rangle~=~\sum_{l=0}^{L}\sum_{i=1,2}C_{il}(v)|X_i, l \rangle$.
We assume each local electronic state undergoes pure dephasing at a rate $\gamma_{pd}$ and associated with jump operator $A_k=|k\rangle \langle k|$, while the collective vibrational mode undergoes thermal emission and absorption processes with rates $\Gamma_{th} (\eta(\omega_{vib})+1)$ and $\Gamma_{th} \eta (\omega_{vib})$, respectively. Here $\eta(\omega)~=~\left(e^{\beta\omega}-1\right )^{-1}$ with $\beta=1/K_B T$ the thermal energy scale. The system is subjected to incoherent pumping of the highest energy exciton $|X_1\rangle$ state, with transition operator $\sigma_{X_1}^{\dag} = |X_1\rangle \langle G|$ and rate $P_{X_1}$. Radiative decay processes from excited vibronic states to the ground are given at rate $\gamma$ and are described by jump operators of the form $\sigma_{vl}=|G,l\rangle \langle F_v|$.  
The Liouvillian of the emitter system in the absence of coupling to any sensor is given by (see \refeq{eq:L0})
\begin{equation}\begin{split}\label{eq:systemME}
	\mathcal{L}_{0}(\hat{\rho}) & = - i \,[H_0,\hat{\rho}]  \,  + \sum_{k={1,2}} \frac{\gamma_{pd}}{2} \mathcal{L}_{A_k}(\hat{\rho}) \\
	& \quad + \, \frac{\Gamma_{th} (\eta(\omega_{vib})+1)}{2} \, \mathcal{L}_{D}(\hat{\rho}) 
	+\frac{\Gamma_{th} \eta (\omega_{vib})}{2} \, \mathcal{L}_{D^{\dag}}(\hat{\rho}) \\
	& \quad+ \, \frac{\gamma}{2} \, \sum_{v=1}^{2L} \sum_{l=1}^{L} \, \mathcal{L}_{\sigma_{vl}}(\hat{\rho}) 
	 \,+ \, \frac{P_{X_1}}{2} \mathcal{L}_{\sigma^{\dag}_{X_1}}(\hat{\rho}).
\end{split}\end{equation}
The sensors, each with bare Hamiltonian $H_m= \omega_m \varsigma^{\dag}_m \varsigma_m $, are assumed to have identical linewidths $\Gamma$ and are coupled with equal strength $\epsilon$ to bare exciton states through the operator $a=\sigma_{X_1}+ \sigma_{X_2}$ such that
$H_{e,m}~=~\epsilon [( \sigma_{X_1}~+~\sigma_{X_2} ) \varsigma^{\dag}_m ~+~(\sigma_{X_1}^{\dag} ~+~\sigma_{X_2}^{\dag}) \varsigma_m] $.
The second term for the Liouvillian superoperator in \refeq{eq:L} then becomes
\begin{equation}\label{eq:sensME}
	\mathcal{L}_{I}(\hat{\rho}) =\sum_{m=1}^{M} \left (\frac{\Gamma}{2} \, \mathcal{L}_{\varsigma_m}(\hat{\rho}) -i \, [H_{m}+H_{e,m},\hat{\rho}] \right ).
\end{equation}
In the right-hand side of Eqs. \eqref{eq:systemME} and \eqref{eq:sensME} $\mathcal{L}_c(O)~ = ~(2cOc^{\dag} - c^{\dag}cO - Oc^{\dag}c)$. We consider bio-inspired parameters for our toy model \cite{OReilly2014,Curutchet2013}. The electronic coupling takes the value $V=92 \, \cm^{-1}$ while the onsite energy difference is $\Delta\alpha=1,042 \, \cm^{-1}$ \cite{OReilly2014} such that the average exciton energy becomes $E =18000 \; \cm^{-1}$ \cite{Curutchet2013} and the exciton energy splitting $\Delta E=1,058.2 \, \cm^{-1}$. The later is comparable with $\omega_{vib}~=~1,111 \, \cm^{-1}$. The thermal energy scale $K_BT=200\, \cm^{-1}$ is on the scale of the coupling strength $g=267.1 \, \cm^{-1}$ but much smaller than $\omega_{vib}$. Hence, a maximum vibrational level of $L=5$ yields converged results. For clarity, in our numerical calculations all wavenumbers are multiplied by $2\pi c$ where $c$ is the speed of light. The electronic pure dephasing is $\gamma_{pd}=[1\, \ps]^{-1}$. We consider an enhanced radiative decay rate $\gamma=[0.5 \,\ns]^{-1}$ and a pumping rate $P_{X_1}=[0.6 \,\ns]^{-1}$. Thermal relaxation is set to $\Gamma_{th}=[4.8 \, \ps]^{-1}$ and equal to the sensor linewidth $\Gamma=[4.8 \, \ps]^{-1}$. 

\begin{figure}[ht]
	\centering
	\includegraphics[width=0.51\textwidth]{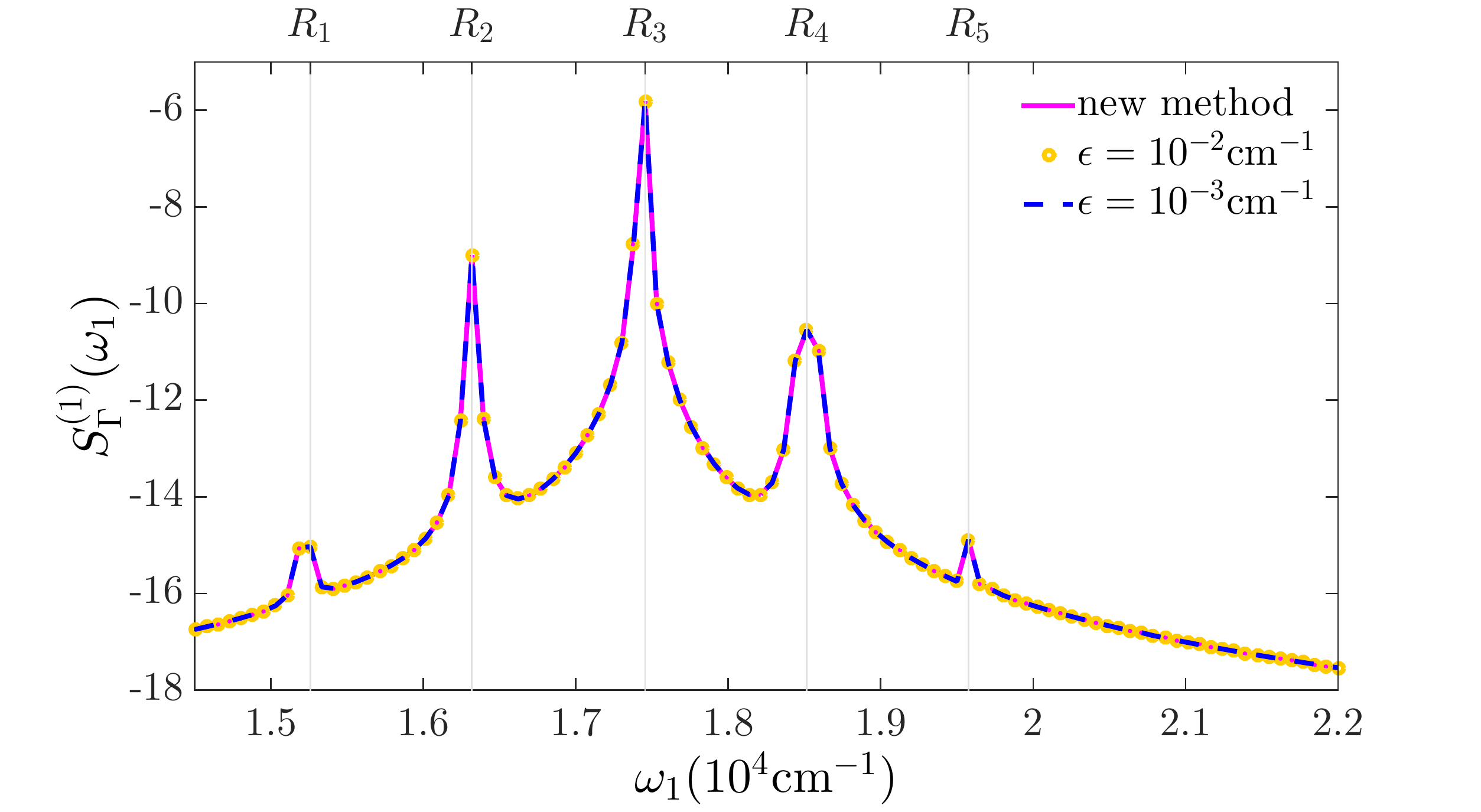} 
	\caption{(Color online) Power spectra $S_\Gamma (\omega_1)$ in log scale versus $\omega_1$ for our vibronic dimer as predicted both by the method proposed in this paper and by the $\epsilon$-dependent sensor method. }
	\label{Fig2}
\end{figure}

\begin{figure}[h]
	\centering
	\includegraphics[width=0.51\textwidth]{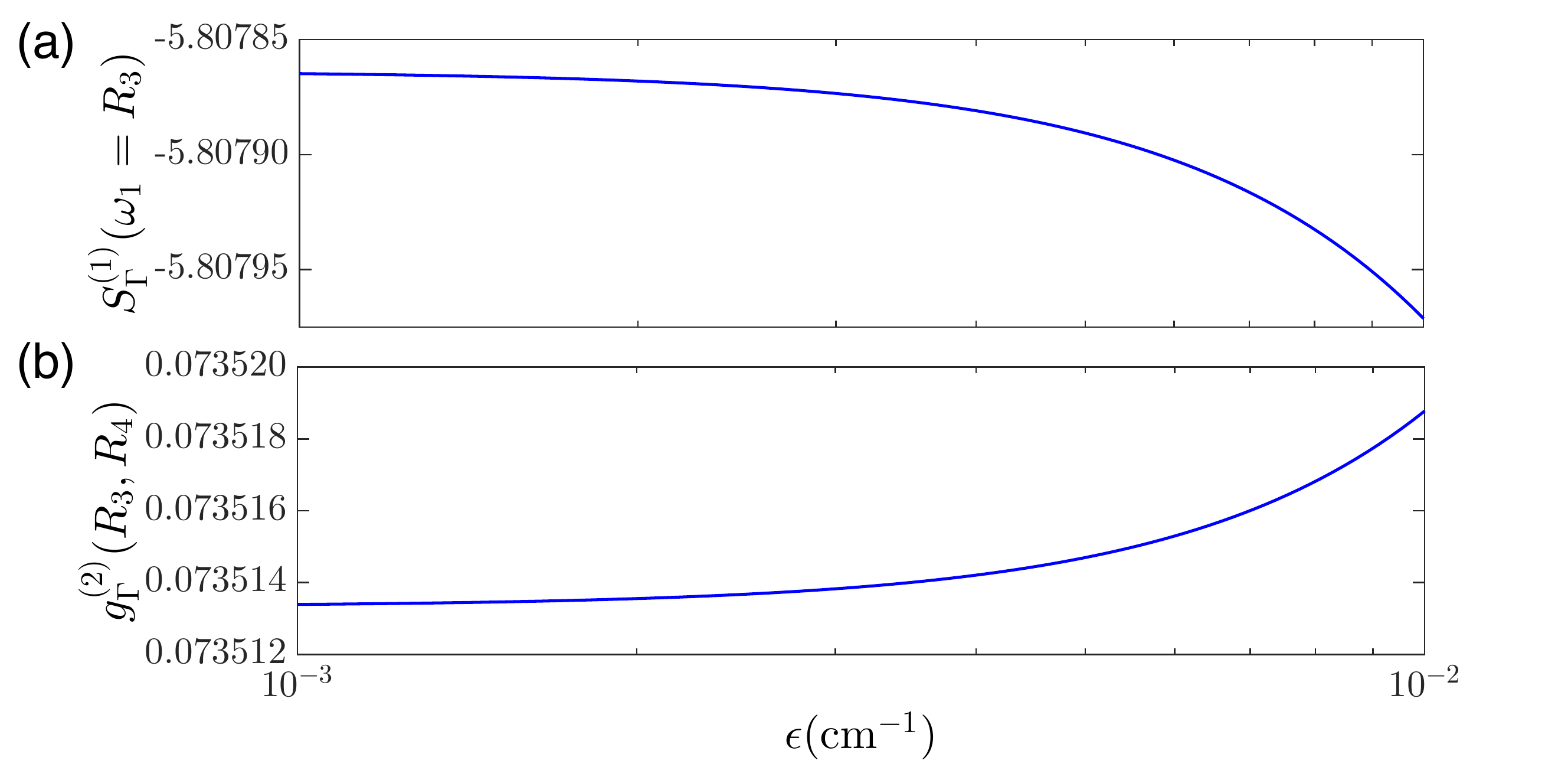} 
	\caption{(Color online) (a) Intensity of power spectrum at a fixed frequency, $S_\Gamma (\omega_1=R_3)$ in log-log scale, and (b) zero-delay time second-order correlation  $g_{\Gamma}^{(2)}(R_4, R_3)$ in semilog scale, as functions of $\epsilon$. Both functions are calculated with the $\epsilon$-dependent method for our vibronic dimer.}
	\label{fig:convergence}
\end{figure}

\reffig{Fig2} presents the power spectra $S_\Gamma (\omega_1)$ for our vibronic dimer as predicted by our approach and by the original method with different values of $\epsilon$ satisfying $\epsilon\ll\sqrt{\Gamma \gamma_{Q}/2}\sim 10^{-1}\,\cm^{-1}$. The highest peak, given at the emission frequency $\omega_1=R_3=17455 \, \cm^{-1}$, captures transitions from the excited vibronic states with the largest amplitude on $|X_2,l\rangle$ to the ground state with the same vibrational quanta $|G,l\rangle$. It also includes transitions from excited states quasi localised on $|X_1,l\rangle \to |G,l+1\rangle$. The peak at $\omega_1=R_4=18515 \, \cm^{-1}$ accounts for transitions from excited vibronic states quasi-localised on $|X_1,l\rangle \to |G,l\rangle$, as well as transitions from states quasi-localised on $|X_2,l\rangle \to |G,l-1\rangle$. The $\epsilon$-dependent method tends to underestimate the spectrum as can be seen in \reffig{fig:convergence} (a), with differences of the order of $\epsilon$. Converged results are obtained for $\epsilon \sim 10^{-3}\textrm{cm}^{-1}$.

\begin{figure}[ht]
	\centering
	\includegraphics[width=0.51\textwidth]{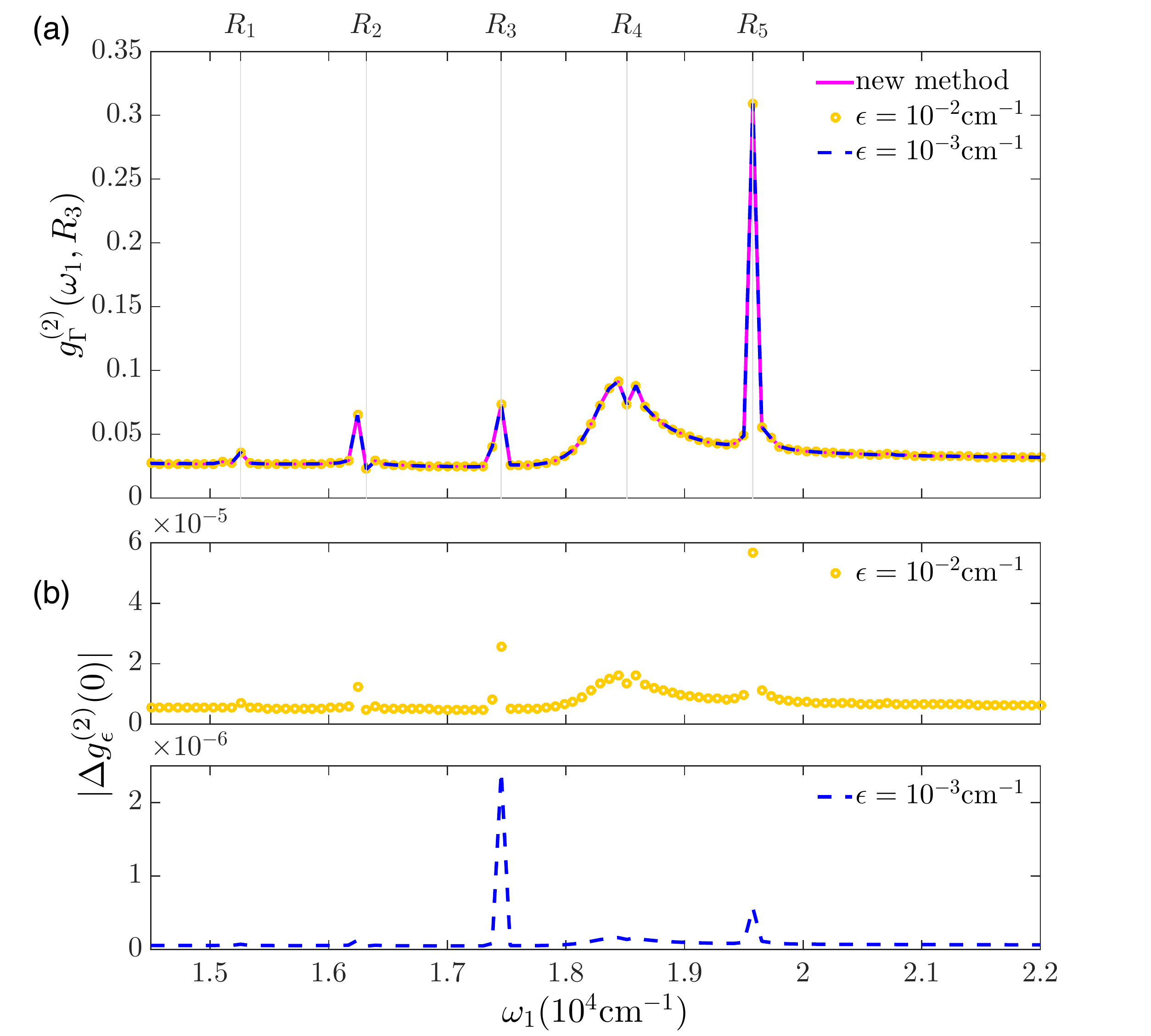} 
	\caption{(Color online) (a) Second order at zero time delay $g_{\Gamma}^{(2)}(\omega_1, R_3)$ versus  $\omega_1$ computed both with the new method and the $\epsilon$-dependent sensor method for different values of $\epsilon$. (b) $|\Delta g^{(2)}_\epsilon(0)|$, the absolute difference value between the predictions of the two methods, versus $\omega_1$ for two values of $\epsilon$.}
	\label{fig:g2R4R3zero}
\end{figure}

The second order correlation function at zero delay time $g_{\Gamma}^{(2)}(\omega_1,\omega_2)$ is shown in \reffig{fig:g2R4R3zero}(a). There we have fixed $\omega_2=R_3$ and scan $\omega_1$ over the domain of frequencies in the power spectrum.  Anti-bunching is observed for the whole frequency regime with larger offsets from zero for the frequency pair $(R_5, R_3)$, indicating transitions corresponding to this pair are weakly correlated. The predictions of the two methods agree up to differences that scale with $\epsilon^2$ as can be seen in \reffig{fig:g2R4R3zero}(b). This figure plots $|\Delta g^{(2)}_\epsilon(0)|$, 
the absolute value of the difference between the values obtained with our approach (solving \refeq{eq:ss_derv}) and the $\epsilon$-dependent method. The later tends to overestimate the second-order photon correlations as can be seen in \reffig{fig:convergence}(b), which plots $g_{\Gamma}^{(2)} (R_4,R_3)$ as function of $\epsilon$.

\begin{figure}[h]
	\centering
	\includegraphics[width=0.51\textwidth]{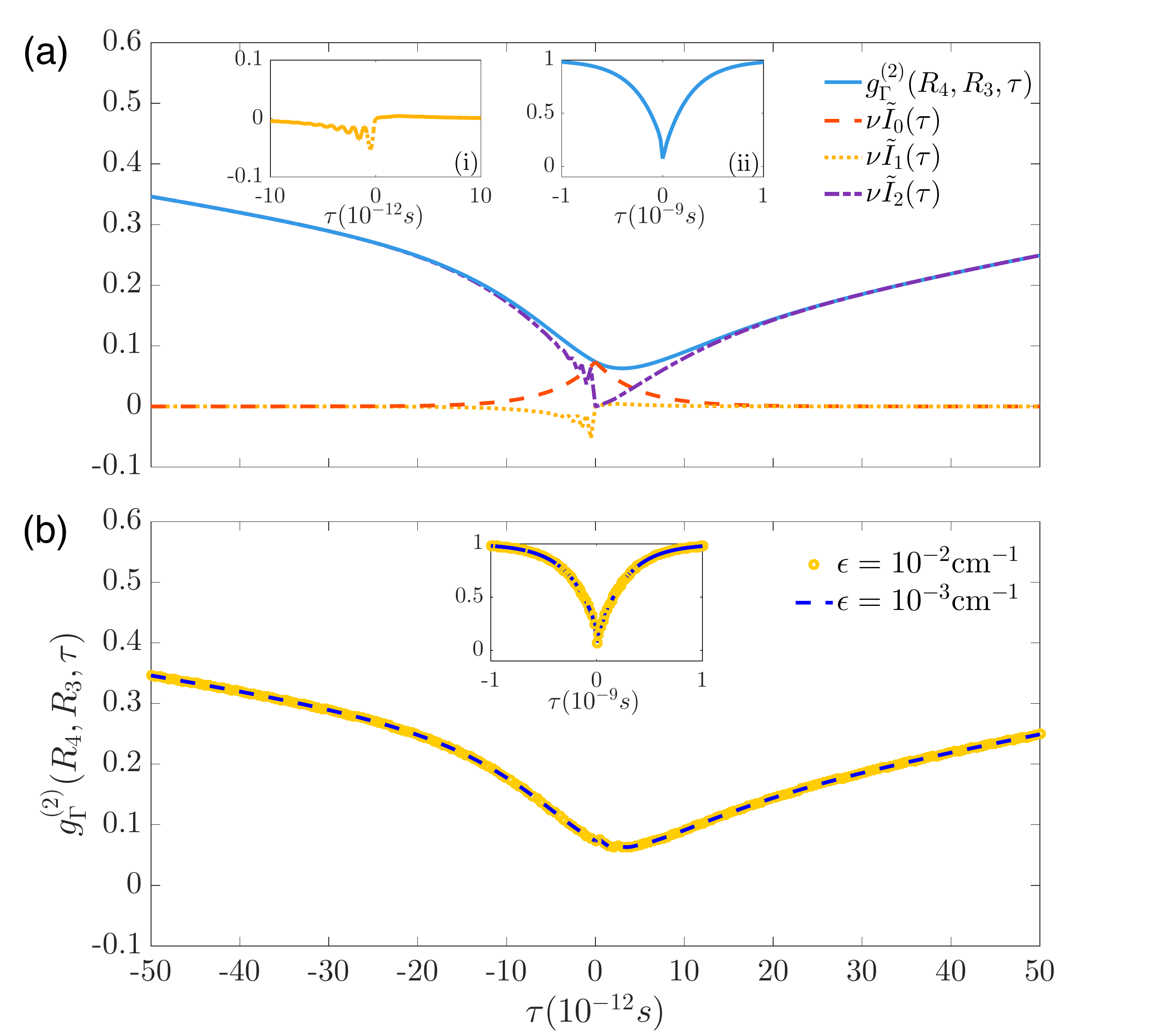} 
	\caption{(Color online) Frequency- and time-resolved correlation function $g_{\Gamma}^{(2)}(R_4, R_3, \tau$) versus $\tau$ predicted with (a) the proposed method and (b) the $\epsilon$-dependent method. Inset $(i)$ in panel (a) shows the short time behaviour of $\nu \tilde{I}_0(\tau)$ as defined in the text. Inset $(ii)$ in (a) and inset in (b) show long time regime of $g_{\Gamma}^{(2)}(\omega_1,\omega_2,\tau)$.}
	\label{fig:g2tR4R3}
\end{figure}

We now turn the attention to the function $g_{\Gamma}^{(2)}(R_4, R_3, \tau)$ depicted in Figs. \ref{fig:g2tR4R3}(a) and (b), which show the correlation between photodetections of the frequency pair $(\omega_1=R_4,\omega_2=R_3)$ as a function of the delay time. We compute this time-resolved correlation in two ways. First, using \refeq{eq:g2taufinal}, we perform the numerical integration for the contributions $\tilde{I}_0(\tau)$,  $\tilde{I}_1(\tau)$ and $\tilde{I}_2(\tau)$ and add them together (\reffig{fig:g2tR4R3}(a)). Second, we use  the $\epsilon$-dependent method (\reffig{fig:g2tR4R3}(b)). The agreement between the predictions of the two methods is evident for both short-time (main panels) and long-time regimes (inset (ii) in \reffig{fig:g2tR4R3}(a) and inset in \reffig{fig:g2tR4R3}(b)). 

The figures highlight the asymmetric behaviour of $g_{\Gamma}^{(2)}(R_4, R_3, \tau)$ with respect to $\tau$, which appears in the time scale of the vibronic decoherence in our model (set mainly by $\Gamma_{th}$). The components $\nu \tilde{I}_k(\tau) \, (k=0,1,2)$ with $\nu=[\ev{n_1}\ev{n_2}]^{-1}$, are also plotted in \reffig{fig:g2tR4R3}(a). As predicted, $\tilde{I}_0(\tau)$ decays exponentially from the initial value set by $g_{\Gamma}^{(2)}(R_4, R_3, \tau=0)$. 
$\tilde{I}_1(\tau)$ is linear in $\tau$ in the short time regime and evolves to take negative values (see \reffig{fig:g2tR4R3}(a) (Inset (i))), reflecting an overdamped oscillation that decays to zero in the long-time regime, in agreement with the behaviours discussed in section \ref{sec:approximations}. The negative values of $\tilde{I_1}(\tau)$ are counteracted by $\tilde{I}_0(\tau)$ and $\tilde{I}_2(\tau)$ such that a physical $g_{\Gamma}^{(2)}(R_4, R_3, \tau)$ is always obtained.  Fig. \ref{fig:g2tR4R3} also shows that the short-time asymmetry in $g_{\Gamma}^{(2)}(R_4, R_3, \tau)$ can be traced back, as expected, to $\tilde{I}_1(\tau)$ and $\tilde{I}_2(\tau)$, indicating that the correlation function is capturing coherent processes in this time scale. Depending on which frequency is probed first, such coherent processes set a different rate for approaching the uncorrelated steady-state emission at large times (see inset (ii) in \reffig{fig:g2tR4R3}(a)).

In summary, we have shown the method here proposed is equivalent to the $\epsilon$-dependent sensor method  to compute frequency-filtered correlation functions. In our method the dependence of correlation functions on $\epsilon$ vanishes algebraically. It therefore avoids both the need to test for convergence for different values of $\epsilon$, and the possible numerical instabilities associated to the smallness of this factor. Identifying \textit{a priori} when the original method will lead to instabilities is difficult, as it is case-dependent.


\section{Conclusion \label{sec:conc}}
We have developed an alternative formulation of the sensor method for the calculation of frequency filtered and time-resolved second-order photon correlations and our main results are summarised by Eqs. (\ref{eq:SMgeneral})-(\ref{eq:general}) and \refeq{eq:g2taufinal}. Our approach, being based on perturbation theory, assumes that the emitter-sensor coupling strength $\epsilon$ is small, but  naturally, the dependence of correlation functions on $\epsilon$ vanishes algebraically. This implies that numerical computation of correlations does not depend explicitly on the value of $\epsilon$. Our method then eliminates the need of evaluating convergence with respect to it, as it is the case when one performs numerical calculations using the original sensor method. Most importantly, the formalism re-defines the problem of computing photon correlations in terms of auxiliary matrices defined in Hilbert space of the emitter only, thereby reducing the dimensionality of the space needed for calculations and, in this way, leading to efficient numerical performance. Provided that, one can relax the quantum regression approach that was used in \cite{DelValle2012} to prove the equivalence between the sensor and integral methods for computing $M$-photon correlations, the relations in Eqs. (\ref{eq:SMgeneral})-(\ref{eq:general}) and \refeq{eq:g2taufinal} apply to a general non-Markovian, non-perturbative open quantum dynamics of the emitter. 

Our proposed method for time-resolved correlations is based on time-dependent perturbation and leads to the expression of $g^{(2)}_{\Gamma_1, \Gamma_2}(\omega_1, \omega_2, \tau)$ in \refeq{eq:g2taufinal} as the sum of three components $\tilde{I}_{0}, \tilde{I}_{1}$ and $\tilde{I}_{2}$, each of which gives insight into the the physical processes dominating the correlations at different time-scales. The trade off is that computation of two of these components requires numerical integration of manageable single and double integrals. The method can be systematically generalised to $M$-photon correlations for zero delay time or when there is delay in only one of the detectors. Its extension to multiple time delays is more elaborated. In that case one can still take the advantage of computing the auxiliary matrix operators given in \refeq{eq:general} but propagate in time without perturbation, thereby combing the advantages of both our approach and the original sensor method. 

To illustrate the agreement between the new approach and the original $\epsilon$-dependent method, we have compared their predictions for the frequency-filtered photon statistics of a toy model that has been inspired in a light-harvesting vibronic dimer. The focus here has been on highlighting the equivalence between the predictions of the two numerical approaches rather than a detailed analysis of the physical insight of the photon-correlations for the system under consideration. We would however like to point out that the results presented here already suggest that frequency-filtered and time-resolved photon-counting statistics can provide a powerful approach to probe coherent contributions to the emission dynamics of biomolecular systems. An in-depth analysis of frequency-filtered photon correlations for the system of interest will be presented in a separate forthcoming manuscript.

\begin{appendices}
\section{Consistency check of the proof of the equivalence between the sensing and the integral methods}\label{app:proof}
In Ref.\cite{DelValle2012} and in its supplemental material it is shown that the sensor method to evaluate $M-$photon correlation functions is identical to the integral method with Lorentzian frequency filter functions for the sensors. Originally, in Ref.\cite{DelValle2012} the normal order for sensor intensity correlations given in \refeq{eq:corr_func} was omitted. This could lead to the confusion that the normal order for sensor operators was, in general, unnecessary. In an Erratum \cite{DelValleErratum} the authors have clarified that their proof assumes normal order all throughout. Since our proposed approach is equivalent to the sensor method, as long as the normal order of the sensor operators is taken into consideration, we have made a consistency check of the proof presented in the supplemental material of Ref.\cite{DelValle2012}.

We begin by considering Eq.(42) in the supplemental material of Ref.\cite{DelValle2012}:
\begin{equation}
\partial_{\tau} \langle n_1(0) n_2(\tau) \rangle= - \Gamma_2 \langle n_1(0) n_2(\tau) \rangle+ 2 \rm{Re} [i\epsilon_2 \langle n_1(0) (\varsigma_2 a^{\dagger})(\tau) \rangle]\, ,
\label{eq:42}
\end{equation}
with $n_j = \varsigma_j^{\dagger} \varsigma_j$ the sensor number operator and $\langle n_1(0) n_2(\tau) \rangle ~\equiv~ \mathrm{Tr}[n_2(\tau) \hat{\rho}_{ss} n_1] $. This equation, which does not consider normal ordering as written, leads to spurious results such as negative values in $g^{(2)}(\omega_1, \omega_2, \tau)$. To see this, we write the steady state density matrix for the joint emitter and sensors as in \refeq{eq:SS_def_two} in our manuscript. In this form the difference between using normally ordered operators and the number operator is evident:
\begin{align}
\varsigma_1 \hat{\rho}_{ss} \varsigma_1^{\dagger} &=  \sum_{j_2,j_2'=0,1}  \hat{\rho}_{1,j_2}^{1,j_2'}  \otimes \ket{j_2} \bra{j_2'} \otimes \ket{0_1} \bra{0_1}  \;. \\
\hat{\rho}_{ss}  \varsigma_1^{\dagger} \varsigma_1 &=   \sum_{j_2,j_2'=0,1} \ket{j_2} \bra{j_2'}  \otimes \left( \hat{\rho}_{1,j_2}^{1,j_2'} \otimes   \ket{1_1} \bra{1_1} \right. \nonumber \\
 &+ \left. \hat{\rho}_{0,j_2}^{1,j_2'}  \otimes  \ket{0_1} \bra{1_1}  \right)  \;. 
\end{align}
First, notice that while these two expressions are different, their traces are identical i.e. $\textrm{Tr}[{n_2 \hat{\rho}_{ss}  n_1}]=\textrm{Tr}[{n_2 \varsigma_1 \hat{\rho}_{ss} \varsigma_1^{\dagger}}]$, which means at $\tau=0$ the normal order for computation of the second-order photon correlation can be waived. However, the difference in these expressions does have an impact for $\tau\ne 0$.
The second expression has the term $\hat{\rho}_{1,j_2}^{1,j_2'} \otimes \ket{1_1} \bra{1_1}$ rather than $\hat{\rho}_{1,j_2}^{1,j_2'} \otimes \ket{0_1} \bra{0_1}$ in the first one; it also contains an additional term $\hat{\rho}_{0,j_2}^{1,j_2'}\otimes\ket{0_1} \bra{1_1} $, which makes the expression not Hermitian (the Hermitian conjugate term with $\ket{1_1} \bra{0_1}$ vanishes due to the action of $n_1$). The impact of this difference becomes clear when we consider $\langle n_1(0) n_2(\tau) \rangle \equiv \mathrm{Tr}[n_2 (\hat{\rho}_{ss} n_1)(\tau)]$. This equation indicates the population of sensor 1 decays exponentially in time with a rate $\Gamma_1$. In terms of derivatives in $\tau$ this means the $\hat{\rho}_{1,j_2}^{1,j_2'}  \otimes \ket{j_2}\bra{j_2'}$ in $(\hat{\rho}_{ss}  \varsigma_1^{\dagger} \varsigma_1)(\tau)$ will acquire an extra factor of $-\Gamma_1$ when compared to those in $( \varsigma_1 \hat{\rho}_{ss}  \varsigma_1^{\dagger})(\tau)$, which is not included in \refeq{eq:42}.  This then disproves \refeq{eq:42}.   

On the other hand, a similar equation for the normally ordered correlation does hold, that is, 
\begin{align}
\partial_{\tau} \langle &\varsigma^{\dagger}_1(0) n_2(\tau) \varsigma_1(0) \rangle = - \Gamma_2 \langle \varsigma^{\dagger}_1(0) n_2(\tau) \varsigma_1(0) \rangle \nonumber \\
&+2 \rm{Re} [i\epsilon_2 \langle \varsigma_1^{\dagger}(0) (\varsigma_2 a^{\dagger})(\tau) \varsigma_1(0) \rangle]+O(\epsilon_1^2,\epsilon_2^2) \;.
\label{eq:42corrected}
\end{align}
The solution of this normally ordered derivative in $\tau$ can be found starting from a vector analogous to $\mathbf{w}'[11,\mu_2\nu_2](\tau)$ given by Eq. (43) in the supplemental material of \cite{DelValle2012} but that contains the terms in normal order:
\begin{align}
\label{eq:wnormalorder}
\tilde{\mathbf{w}}[11,\mu_2\nu_2](\tau)  &= \begin{bmatrix}
          \ev{ \varsigma^{\dagger}_1 (\varsigma_2^{\dagger,\mu_2} \varsigma_2^{\nu_2})(\tau) \varsigma_1 }  \\
		\ev{ \varsigma^{\dagger}_1 (\varsigma_2^{\dagger,\mu_2} \varsigma_2^{\nu_2} a)(\tau) \varsigma_1 } \\
          \ev{ \varsigma^{\dagger}_1 (\varsigma_2^{\dagger,\mu_2} \varsigma_2^{\nu_2} a^{\dagger})(\tau) \varsigma_1 } \\
          \ev{ \varsigma^{\dagger}_1 (\varsigma_2^{\dagger,\mu_2} \varsigma_2^{\nu_2} a^{\dagger}a)(\tau) \varsigma_1 } \\
          \vdots 
         \end{bmatrix} \;.
\end{align}
The time derivatives of the elements in  $\tilde{\mathbf{w}}[11,\mu_2\nu_2](\tau)$ are of the form
\begin{align}
\partial_{\tau} \ev{ \varsigma^{\dagger}_1 (\varsigma_2^{\dagger\mu_2} \varsigma_2^{\nu_2} a^{\dagger \nu}a^{\nu'}  )(\tau) \varsigma_1 }  = \mathrm{Tr}\{ (\varsigma_2^{\dagger,\mu_2} \varsigma_2^{\nu_2} a^{\dagger \nu}a^{\nu'} )(\tau)  \mathcal{L}(\varsigma_1 \rho_{ss} \varsigma^{\dagger}_1)  \}\;,
\label{eq:PDw}
\end{align}
where the Liouvillian is defined as in \refeq{eq:L} in this manuscript. In particular, we are interested in obtaining an equation when $\mu_2=0$ and $\nu_2=1$.

Following the formalism either of the supplemental material of \cite{DelValle2012} or a time-dependent perturbation approach as we propose in our manuscript, one can show that, in the limit $\ev{n_{1(2)}} \ll 1$, the solution for the normally ordered correlation is formally identical to Eq. (44) in Ref. \cite{DelValle2012} (supplemental material): 
\begin{align}
\partial_{\tau} \tilde{\mathbf{w}}[11,01](\tau) = &[\mathbf{M}-(i\omega_2 + \Gamma_2/2) \openone ] \tilde{\mathbf{w}}[11,01](\tau) \nonumber \\
&- i\epsilon_2 \mathcal{T}_{-}\tilde{\mathbf{w}}[11,00](\tau), 
\end{align}
where $\mathbf{M}$ is the matrix that rules the dynamical evolution of the emitter. This means the equations governing the normally ordered vector $\tilde{\mathbf{w}}[11,\mu_2\nu_2](\tau)$ (\refeq{eq:wnormalorder}) are exactly the same as those presented in the proof given in \cite{DelValle2012} (supplemental material), in agreement with the clarification stated in the Erratum \cite{DelValleErratum}. 

\section{Numerical procedure to compute zero delay time correlations of order $M>2$.}\label{app:M}
Starting with the steady state for the joint emitter and $M$ sensors written as in \refeq{eq:SS_def_ex},  the $M$th order photon correlation at $\tau=0$ depends on the rescaled matrix $\tilde{\hat{\rho}}_{1\ldots1}^{1\ldots1}~=~\langle 1_1, \dots, 1_M | \tilde{\hat{\rho}}_{ss} |1_1, \ldots,1_M \rangle$. To find this matrix we solve $\mathcal{L} (\hat{\rho}_{ss}) = 0$ which, in analogy to Eq.(14), can be re-written as 
\begin{equation}\begin{split}
\mathcal{L} (\hat{\rho}_{ss)}&= \sum_{j_1, j_1'\dots j_M, j_M'} \hat{B}_{j_1 \ldots j_m\dots j_M}^{j_1' \ldots j_m'\dots j_M'} \otimes 
\ket{j_1} \bra{j_1'}\ldots \otimes \ket{j_m} \bra{j_m'}\\ 
& \quad \dots \otimes \ket{j_M} \bra{j_M'}=0
\end{split}\end{equation}
We derive the set of equations satisfying $\hat{B}_{j_1 \ldots j_m\dots j_M}^{j_1' \ldots j_m'\dots j_M'}=\mathbf{0}$ with the approximation of ignoring down coupling as discussed in the main text.  This leads to a hierarchy satisfied by the matrices $\tilde{\hat{\rho}}_{j_1 \ldots j_m\dots j_M}^{j_1' \ldots j_m'\dots j_M'}$. Careful inspection of the sets in \refeq{eqs:SS_one_sensor} and \refeq{eqs:SS_two_sensor} allows to identify the pattern for such a set. Let us define $J = j_1+j_1'+\ldots+j_m+j_m'\dots + j_M+j_M'$ the sum of all matrix indexes. Notice that for $J=0$ the solution is simply given by $\mathcal{L}_0 \left (\tilde{\hat{\rho}}_{0,\ldots,0}^{0,\ldots,0}\right ) \sim 0$. In general, the form for the left hand side terms for each equation will be given by
\begin{align}
\label{eq:LHS}
\left[\mathcal{L}_0-\sum_{m=1}^M \left \{(j_m+j'_m)\Gamma_m/2+(j_m-j'_m)i\omega_m)\right \} \right] \tilde{\hat{\rho}}_{j_1 \ldots j_m\dots j_M}^{j_1' \ldots j_m'\dots j_M'}.
\end{align}
This term is simply down to the evolution under the Liouvillian of the emitter and the decay and phase evolution of the sensors. Each matrix with $J\ge 1$ is coupled only to matrices with $J-1$.  Hence, the solution of the $J=2M$ matrix $\tilde{\hat{\rho}}_{1,\ldots,1}^{1,\ldots,1}$ involves only matrices with $J=2M-1$ and so on (cf. \refeq{eq:ss_derv}j). The total number of tier-below matrices required equals $J$ and each of these matrices differs only in one index $j_m$ or $j_m'$ which will be $0$ rather than unity. Let us call this tier-below matrices $\tilde{\hat{\rho}}_{\ell_1 \ldots \ell_m \dots \ell_M}^{\ell_1' \dots \ell_m' \ldots \ell_M'}$. The matrix that differs in the $m$th component such that $j_m = 1$ and $\ell_m = 0$, with all others equal, will add a term of the form $i {a}_m \tilde{\hat{\rho}}_{j_1 \ldots j_m=0\ldots j_M}^{j_1' \ldots j_m' \ldots j_M'}$. Likewise if $j'_m = 1$ but $\ell'_m = 0$  and $\ell'_y = j'_y$ and $\ell_y = j_y$ for $y \neq m$, we have a contribution of the form $-i \tilde{\hat{\rho}}_{j_1 \ldots j_m\ldots j_M}^{j_1' \ldots j_m'=0 \ldots j_M'} {a}^{\dagger}_m$. Therefore, the right-hand side term, to which \refeq{eq:LHS} is approximated to, will be of the form:
\begin{align}
i \sum_{m=1}^M \left [ \delta_{j_m,1}{a}_m \tilde{\hat{\rho}}_{j_1 \ldots j_m(1 -\delta_{j_m,1})\ldots j_M}^{j_1' \ldots j_m' \ldots j_M'}
- \delta_{j_m',1} \tilde{\hat{\rho}}_{j_1 \ldots j_m\ldots j_M}^{j_1' \ldots j_m' (1-\delta_{j_m',1}) \ldots j_M'} a_m^{\dagger}\right ].
\end{align}
Here $\delta_{u,v}$ is the Kronecker delta function, equal to zero if $u \neq v$ or unity if $u=v$. 

 \end{appendices}
 
 \acknowledgements 
We thank Elena del Valle, Juan Camilo L\'opez Carre\~no, Carlos S\'anchez Mu\~noz and Fabrice P. Laussy for discussions. Financial support from the Engineering and Physical Sciences Research Council (EPSRC UK) and from the EU FP7 Project PAPETS (GA 323901) is gratefully acknowledged.


%

\end{document}